\documentclass[twocolumn]{aastex631}

\usepackage{CJK}
\usepackage{times}
\usepackage{amsmath}
\usepackage{graphicx}
\usepackage{hyperref}
\usepackage{gensymb}
\usepackage{upgreek}
\usepackage{natbib}
\usepackage{subfigure}
\usepackage{multirow}
\usepackage{comment}
\usepackage{mathtools}
\usepackage{mathrsfs}
\usepackage{fontenc}
\usepackage{color}
\usepackage{url}
\usepackage{pifont}

\def\lx{\textit{$L_{\rm X}$}}
\def\nh{\textit{$N_{\rm H}$}}
\newcommand{\sersic}{S\'{e}rsic}
\newcommand{\msersic}{$m_{\rm S\acute{e}rsic}$}

\begin{document}
\begin{CJK*}{UTF8}{gbsn}

\title{AGNs and Host Galaxies in COSMOS-Web. I. NIRCam Images, PSF Models and Initial Results on X-ray-selected Broad-line AGNs at $0.35\lesssim z \lesssim 3.5$}

\author[0000-0001-5105-2837]{Ming-Yang Zhuang (庄明阳)}
\email{mingyang@illinois.edu}
\affiliation{Department of Astronomy, University of Illinois at Urbana-Champaign, Urbana, IL 61801, USA}

\author[0000-0002-1605-915X]{Junyao Li}
\affiliation{Department of Astronomy, University of Illinois at Urbana-Champaign, Urbana, IL 61801, USA}

\author[0000-0003-1659-7035]{Yue Shen}
\affiliation{Department of Astronomy, University of Illinois at Urbana-Champaign, Urbana, IL 61801, USA}
\affiliation{National Center for Supercomputing Applications, University of Illinois at Urbana-Champaign, Urbana, IL 61801, USA}

\begin{abstract}
We present detailed and comprehensive data reduction and point-spread-function (PSF) model construction for all public JWST NIRCam imaging data from the COSMOS-Web treasury program (up to June 2023, totaling 0.28 ${\rm deg}^2$). We show that the NIRCam PSF has significant short-timescale temporal variations and random spatial variations in all four filters (F115W, F150W, F277W, and F444W). Combining NIRCam with archival HST imaging, we perform multiwavelength AGN+host image decomposition to study the properties of 143 X-ray-selected ($L_{\rm bol}=10^{43.6-47.2}$ erg\,s$^{-1}$) broad-line AGNs at $0.35\lesssim z \lesssim 3.5$. Leveraging the superb resolution, wavelength coverage, and sensitivity of NIRCam, we successfully detect host stellar emission after decomposing the central AGN point source in 142 objects. $\sim 2/3$ AGNs are in star-forming galaxies based on the UVJ diagram, suggesting no instantaneous negative AGN feedback. X-ray-selected broad-line AGN hosts follow a similar stellar mass-size relation as inactive galaxies, albeit with slightly smaller galaxy sizes. We find that although major mergers are rare ($\sim7$--22\%) among the sample, more subtle non-axisymmetric features from stellar bars, spiral arms, and minor mergers are ubiquitous, highlighting the importance of secular processes and minor mergers in triggering AGN activity. For a subsample of 30 AGNs at $1<z<2.5$ with black hole mass measurements from single epoch spectra, they follow a similar black hole mass-stellar mass relation as local inactive early-type galaxies but reside preferentially near the upper envelope of nearby AGNs. We caution that selection biases and intrinsic differences of AGN populations at different redshifts may significantly affect their location on the black hole mass-stellar mass plane.

\end{abstract}

\keywords{}

\section{Introduction} \label{Sec1}
The discovery of tight correlations between the masses of the supermassive black holes (BHs) and the properties of their host galaxies (such as stellar velocity dispersion and bulge/total stellar mass) in the local Universe suggests that BHs and galaxies may coevolve with each other \citep[e.g.,][]{Magorrian+1998AJ, Gebhardt+2000ApJ, Kormendy&Ho2013ARA&A}. Popular scenarios propose that the feedback from active galactic nuclei (AGNs) plays an important role in regulating the growth of the BH and its host galaxy by injecting energy and momentum into their environment \citep[e.g.,][]{McNamara&Nulsen2007ARA&A, Hopkins+2008ApJS, King&Pounds2015ARA&A}. However, the details of these feedback processes and their impact on host galaxies are still being debated. Investigating when and how these correlations are established, as well as obtaining robust properties of AGN host galaxies, such as morphology, structure, environment, and stellar population, is key to understanding galaxy and BH evolution.

The close track of the cosmic BH accretion history to the cosmic star formation history suggests that star formation and black hole growth are closely connected across cosmic time \citep[e.g.,][]{Boyle&Terlevich1998MNRAS, Silverman+2008ApJ, Kormendy&Ho2013ARA&A, Madau&Dickinson2014ARA&A}, at least in the global sense. At cosmic noon ($z\approx2$), when star formation and BH accretion reach their peak epoch, we would expect stronger AGN feedback at play, making it the ideal epoch to investigate the cosmic evolution of the BH-galaxy scaling relations. However, previous studies of BH mass ($M_{\rm BH}$)--host stellar mass ($M_*$) relations of AGNs during this epoch tend to produce contradictory results, with AGNs lying above, on or below the location relation \citep[e.g.,][]{Borys+2005ApJ, Alexander+2008AJ, Jahnke+2009ApJ, Merloni+2010ApJ, Sun+2015ApJ, Suh+2020ApJ, Ding+2020ApJ, Zhang2023ApJ}. Various factors may account for the large, apparent discrepancies among different works, including measurement uncertainties and limited dynamical ranges of $M_{\rm BH}$ and $M_*$, small sample statistics, and selection biases \citep[e.g.,][]{Lauer+2007ApJ, Shen&Kelly2010ApJ, Schulze&Wisotzki2011A&A, Shankar+2016MNRAS, Li2021mass}. 

On the other hand, no consensus has been reached in terms of whether major merger is the primary triggering mechanism of AGN activity. Based on different AGN samples with imaging data of various depths and resolutions, some find a low merger fraction ($\lesssim20$\%), while others find merger fractions that can be as high as $40-50$\%, with tentative evidence that the merger fraction increases as AGN luminosity increases \citep[e.g.,][]{Treister+2012ApJ, Villforth+2014MNRAS, Hong+2015ApJ, Villforth+2017MNRAS, Ellison+2019MNRAS, Gao+2020A&A, Marian+2020ApJ, Kim+2021ApJS, Tang+2023arXiv}. Meanwhile, it is anticipated that less violent minor mergers and secular processes driven by stellar bars and spiral arms can also deliver gas inflow and feed BH accretion \citep[e.g.,][]{Kormendy&Kennicutt2004ARA&A, Kaviraj+2014MNRAS, Kim&Kim2014MNRAS, Shu2016ARA&A}. 

{When two galaxies merge, a pair of massive BHs may form, whose subsequent evolution may lead to the formation of a bound supermassive black hole binary. If both BHs are active, a dual AGN system may be witnessed.} Alternatively, off-nucleus AGNs may be signpost of the inspiraling phase of two merging galaxies where only one BH is active, or a recoiled BH from gravitational kickout of BH binary coalescence \citep[e.g.,][]{Baker+2006ApJ, Barth+2008ApJ, Campanelli+2007PhRvL, Comerford&Greene2014ApJ}. However, due to the difficulties of robustly identifying these systems observationally, especially at close separations, the sample size of previous works is small and mainly limited to low-redshift ($z<1$) AGNs, or to luminous quasars at $z>1$ \citep{Liu+2011, Comerford+2015ApJ, Shen+2019ApJ,Shen+2023}.

All these science cases above require measurements of AGN host galaxy properties, in particular, properties of the host stellar content, as functions of redshift, AGN luminosity, and black hole mass. At $z\approx2$, rest-frame optical coverage of the Hubble Space Telescope (HST) has limited the study of host structural properties, i.e., F160W corresponds to rest-frame $\sim5300$\AA\ at $z\approx2$. Wavelength coverage of up to 5 \micron\ provided by the James Webb Space Telescope \citep[JWST;][]{Gardner2006SSRv} Near Infrared Camera (NIRCam) significantly improves host galaxy studies at $z\approx 2$ and beyond.  

In this paper, we make use of JWST NIRCam imaging data from the COSMOS-Web treasury program \citep[PIs: Kartaltepe \& Casey, ID=1727]{COSMOS-Web}, targeting at the Cosmic Evolution Survey \citep[COSMOS;][]{Scoville+2007ApJS} field. The unprecedented sensitivity, wavelength coverage (up to $\sim5$ \micron) and resolution (0\farcs05--0\farcs16 from the F115W filter to F444W filter) of NIRCam imaging makes it possible to study morphological, structural and stellar  properties of $z\approx2$ AGNs and galaxies at sub-kpc scale at rest-frame optical and near-infrared (NIR) for the first time.

Given the tremendous amount of JWST data in COSMOS-Web and broad science applications of these data, we are conducting a series of initial studies focusing on AGNs and their host galaxies. In Paper I (this paper), we focus on JWST NIRcam data reduction, PSF model construction and present initial results on an X-ray-selected broad-line AGN sample at $0.35\lesssim z \lesssim 3.5$; in Paper II, we will search for candidate offset and dual AGNs \citep[e.g.,][]{Li+2023_CID42}, and investigate the connection between mergers and AGN activity; in Paper III, we will compare host properties with obscured AGNs and inactive galaxies to study the role of AGN in the general galaxy population; in Paper IV, we will perform a detailed study of the substructures of AGN host galaxies, such as bulges, bars, and spiral arms. Expanding on literature AGN samples in the COSMOS field, ultimately we plan to perform our own photometric classification upon JWST imaging data and select AGN samples to much fainter magnitudes and higher redshifts than previous samples, thus extending the dynamical range of our AGN and host galaxy samples in COSMOS-Web.

This paper is structured as follows. Section~\ref{Sec2} describes the data and data reduction. Section~\ref{Sec3} presents point-spread-function (PSF) construction procedures and properties of NIRCam PSFs in COSMOS-Web. Section~\ref{Sec4} describes the X-ray broad-line AGN sample, followed by derivation of host properties including host structural parameters from AGN-host image decomposition and stellar properties from fits to their spectral energy distributions (SEDs) in Section~\ref{Sec5}. Section~\ref{Sec6} discusses the properties of AGN host galaxies, AGN host offset, and the $M_{\rm BH}-M_*$ relation. Our main conclusions are summarized in Section~\ref{Sec7}. We adopt a flat $\Lambda$CDM cosmology with $H_0$=70 km s$^{-1}$ Mpc$^{-1}$, $\Omega_{\rm m}$ = 0.3, and $\Omega_{\rm \Lambda}= 0.7$, and a \citet{Chabrier2003PASP} initial mass function (IMF) for stellar population analysis. The reduced NIRCam imaging data as well as PSF models are publicly available at \url{https://ariel.astro.illinois.edu/cosmos_web/}.

\section{Data} \label{Sec2}
COSMOS-Web \citep[PIs: Kartaltepe \& Casey, ID=1727]{COSMOS-Web} is a 255 hour JWST treasury program targeting the central area of the COSMOS field. It covers a contiguous area of 0.54 deg$^2$ with NIRCam imaging in four filters: two short-wavelength filters (SW): F115W and F150W, and two long-wavelength filters (LW): F277W and F444W. At the same time, a 0.19 deg$^2$ area of MIRI imaging with a single filter (F770W) is observed in parallel with NIRCam observations. At the time of writing, 80 visits have been successfully executed (January 2023 and April-May 2023), covering roughly half of the entire COSMOS-Web field. The remaining visits are scheduled in December 2023 and January 2024. Moreover, the entire COSMOS-Web field is covered by HST Advanced Camera for Surveys (ACS) Wide Field Channel (WFC) $I$ band (F814W) observations \citep{Koekemoer+2007ApJS}, extending the wavelength coverage to rest-frame ultraviolet (UV)/optical for relatively low redshift sources ($\sim$0.4--0.8 \micron\ at $z\lesssim1$). 

In this paper, we make use of NIRCam imaging data of the COSMOS-Web survey and HST F814W mosaics (\texttt{v2.0})\footnote{\url{https://irsa.ipac.caltech.edu/data/COSMOS/images/acs_mosaic_2.0/}} from the COSMOS survey \citep{Koekemoer+2007ApJS, Massey+2010MNRAS}. 

\begin{figure}[t]
\centering
\includegraphics[width=0.5\textwidth]{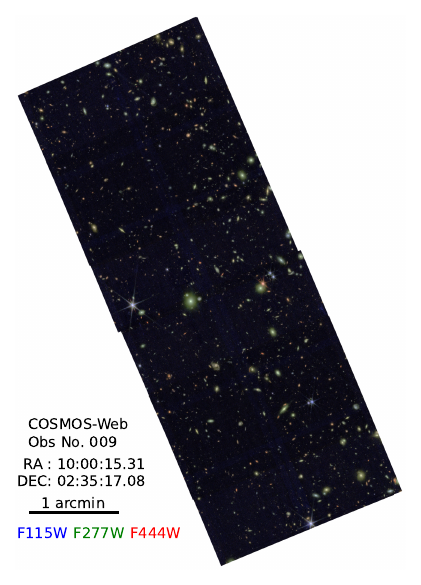}
\caption{Pseudo-color image of the COSMOS-Web observation number 009 constructed from the F444W (red), F150W (green), and F115W (blue) filter images. The scale bar indicates one arcminute. Up direction indicates north and left direction indicates east.}
\label{fig1}
\end{figure}

\subsection{Data Reduction}
We retrieve uncalibrated NIRCam raw data of 80 visits from MAST\footnote{\url{https://mast.stsci.edu/}}. We reduce the data using version 1.10.2 of the \texttt{jwst}\footnote{\url{https://jwst-pipeline.readthedocs.io/en/latest/}} pipeline with the Calibration Reference Data System (CRDS) version of 11.17.0 (context file \texttt{jwst\_1089.pmap}) and custom steps for NIRCam image data reduction. By the time of writing, new pipeline version (1.11.2) and CRDS context file (1097) are available, but none of the updates affects the reduction and calibration of the NIRCam imaging data. All the JWST NIRCam data used in this paper can be found in MAST: \dataset[10.17909/0ktr-qn45]{http://dx.doi.org/10.17909/0ktr-qn45}.
Our detailed reduction procedures are as follows:
\begin{itemize}
    \item We first reduce uncalibrated raw images of individual exposures using the Stage 1 pipeline \texttt{Detector1Pipeline}, which applies basic detection-level corrections. We adopt the default configuration with the exception of three parameters of the \texttt{Jump Detection} step. We turn on \texttt{expand\_large\_events} to flag large cosmic ray events and fine tune \texttt{min\_jump\_area=10} and \texttt{sat\_required\_snowball=False} for better identification and flagging snowballs. We then subtract $1/f$ noise (horizontal and vertical patterns) in the output countrate (slope) images of Stage 1 pipeline using scripts developed by the Cosmic Evolution Early Release Science Survey (CEERS; ERS 1345, PI: Steven Finkelstein) team \citep{Bagley+2023ApJ}.
    
    \item We run Stage 2 pipeline \texttt{Image2Pipeline} to the countrate images with default parameters. This step includes wcs assignment, flat-fielding, and photometric calibration and returns fully calibrated individual exposures. We skip the \texttt{Skymatch} step in Stage 3 pipeline and subtract two-dimensional background after masking sources and bad pixels in individual exposures using \texttt{SExtractorBackground} with a box size of 50 pixels\footnote{The native detector pixelscales are 0\farcs031 for SW channels and 0\farcs063 for LW channels.} implemented in \texttt{photutils} \citep{photutils}. Masks are generated with dilated (using a circular footprint with a radius of 5 pixels) segmentation maps of sources with at least 10 connecting pixels and fluxes above $2\times$pixel-wise standard deviation in convolved images (gaussian kernel with full-width-at-half-maximum (FWHM) $=3$ pixels). We note that our adopted box size may oversubtract faint, diffuse emission of very large objects {(size $\gg50$ pixels)}. 
    
    \item We then subtract the ``wisp'' features caused by scattered light coming off-axis and bouncing off the top secondary mirror strut\footnote{\url{https://jwst-docs.stsci.edu/jwst-near-infrared-camera/nircam-features-and-caveats/nircam-claws-and-wisps}}. Wisps are stationary features present in the same detector locations in all exposures with variable brightness. They are most prominent in the B4 detector, with fainter features in A3, A4, and B3 detectors. For COSMOS-Web observations, wisps affect F115W and F150W images. We create wisp templates by median-stacking the source emission masked-images (product from the previous step). By dividing the observations into two groups: observations in January 2023 and observations in April/May 2023, we find that wisps do not have strong time variations. Therefore, the final templates are based on 320 (4 dithers $\times$ 80 observations) individual exposures for each detector and each filter (see Figure~\ref{figa} for templates in the F150W filter). Wisps in the F115W filter have similar structures but much fainter brightness compared to those in the F150W filter. As we have subtracted the background in the previous step, our wisp templates retain mainly the filamentary features while eliminating the diffuse component compared to the templates provided by the NIRCam team\footnote{\url{https://stsci.box.com/s/1bymvf1lkrqbdn9rnkluzqk30e8o2bne}} and \citet{Robotham+2023arXiv}. We obtain the scaling factor multiplied to the template to subtract the wisps by fitting the template to the source emission-masked images using the \texttt{scipy.optimize.leastsq} function. To quantify the significance of wisps in each exposure, we select a 5 $\times$ 5 pixel window targeted on the brightest wisp feature in each detector and calculate the median ratio within the window between the wisps to be subtracted (template times scaling factor) and the background noise ($\sigma$ from the error extension of calibrated images). The distributions of the wisp significance are shown in Figure~\ref{figb}. A3, A4, and B3 imaged with the F115W filter have wisp significance $<0.1\sigma$, suggesting negligible effects of wisps on these detectors. Therefore, we only perform wisp subtraction in the four detectors imaged with the F150W filter and B4 imaged with the F115W filter.

    \item We also find ``claw'' features in six observations taken in January 2023 (Observation number 043--048). Claws are artifacts that occasionally appear due to scattered light from extremely bright stars located in a specific susceptibility region that is very far from the field-of-view of NIRCam. For COSMOS-Web observations, we find claws in the B1 and B2 detectors imaged with the F150W filter (Figure~\ref{figc}) instead of A1, A2, and B4 detectors reported in JWST documentation\footnote{\url{https://jwst-docs.stsci.edu/jwst-near-infrared-camera/nircam-features-and-caveats/nircam-claws-and-wisps}} and with slightly different appearances (seem to be flipped). This is likely due to the differences in the location of the bright star in the susceptibility region and the position angle of the telescope. As claws move from observations to observations (much smaller movement among four dither exposures within a visit), we manually mask the pixels that are affected for each exposure. These features are at the level of $\sim0.8$ times the background noise $\sigma$ in individual exposures, comparable to the level of wisps in A3 detector in the F150W filter. Note that both wisps and claws are more prominent with respect to the background in smoothed images used for source detection and may be confused as sources with low surface brightness.

    \item Finally, we run Stage 3 pipeline \texttt{Image3Pipeline} to produce a single mosaic for each filter by combining all the calibrated images (all dithers and detectors). We turn on \texttt{tweakreg} step for astrometry correction, \texttt{outlier\_detection} step for outlier pixel rejection (bad or cosmic-ray affected pixels), and \texttt{resample} step for image resample and combination. We correct and tie the astrometry of NIRCam images to that of the COSMOS2020 ``Farmer'' catalog \citep{COSMOS2020}, whose astrometry was tied to Gaia \citep{GAIA_DR1}. We use the updated version \texttt{COSMOS2020\_FARMER\_R1\_v2.2\_p3} from CDS VizieR \citep{vizier:J/ApJS/258/11} and adopt coordinates from model fitting (RAmdeg and DEmdeg) for sources with reliable model fitting results and small drift from detection (\texttt{FModel=0}). We adopt the default configurations for \texttt{outlier\_detection} except for setting \texttt{maskpt=0.5}, which is the fraction of maximum weight to use as the lower limit for valid data. We find that a significant amount of good pixels have weight below 0.7 (default) times the maximum value of the weight map. If not modified, outlier pixels with relatively small weight would not be identified and remain in the final mosaic. For the \texttt{resample} step, we adopt common output image shape (\texttt{output\_shape}), reference pixel position (\texttt{crpix}) and coordinate (\texttt{crval}), pixel scale (\texttt{pixel\_scale=0\farcs03 pixel$^{-1}$}), and input pixel ``shrunk'' fraction (\texttt{pixfrac=0.8}) for all four filters associated with the same observation. All the other parameters are kept as their default values.
    
\end{itemize}

Figure~\ref{fig1} shows an example pseudo-color mosaic of COSMOS-Web NIRCam imaging of observation 009. We also compare positions of bright sources (signal-to-noise ratio, SNR $>50$; containing both resolved and unresolved sources) in our NIRCam mosaics with those in the COSMOS2020 catalog. The agreement with the COSMOS2020 catalog is quite good, with {median differences and standard deviations of coordinates between all sources in the COSMOS2020 catalog and their counterparts in NIRCam images:} $\Delta {\rm RA}=-1\pm6$, $-3\pm4$, $-2\pm3$, $-2\pm3$ mas, $\Delta {\rm DEC}=4\pm4$, $3\pm4$, $1\pm3$, $1\pm3$ mas among 80 visits for the F115W, F150W, F277W, and F444W filters, respectively. The standard deviations of $\Delta {\rm RA}$ and $\Delta {\rm DEC}$ are larger for sources in individual visits (30--40 mas). Wavelength-dependent galaxy morphology, such as more frequent presence of off-nucleus star-forming clumps \citep[e.g.,][]{Elmegreen+2007ApJ, Guo+2015ApJ}, may contribute to the scatter. 

\begin{figure}[t]
\centering
\includegraphics[width=0.5\textwidth]{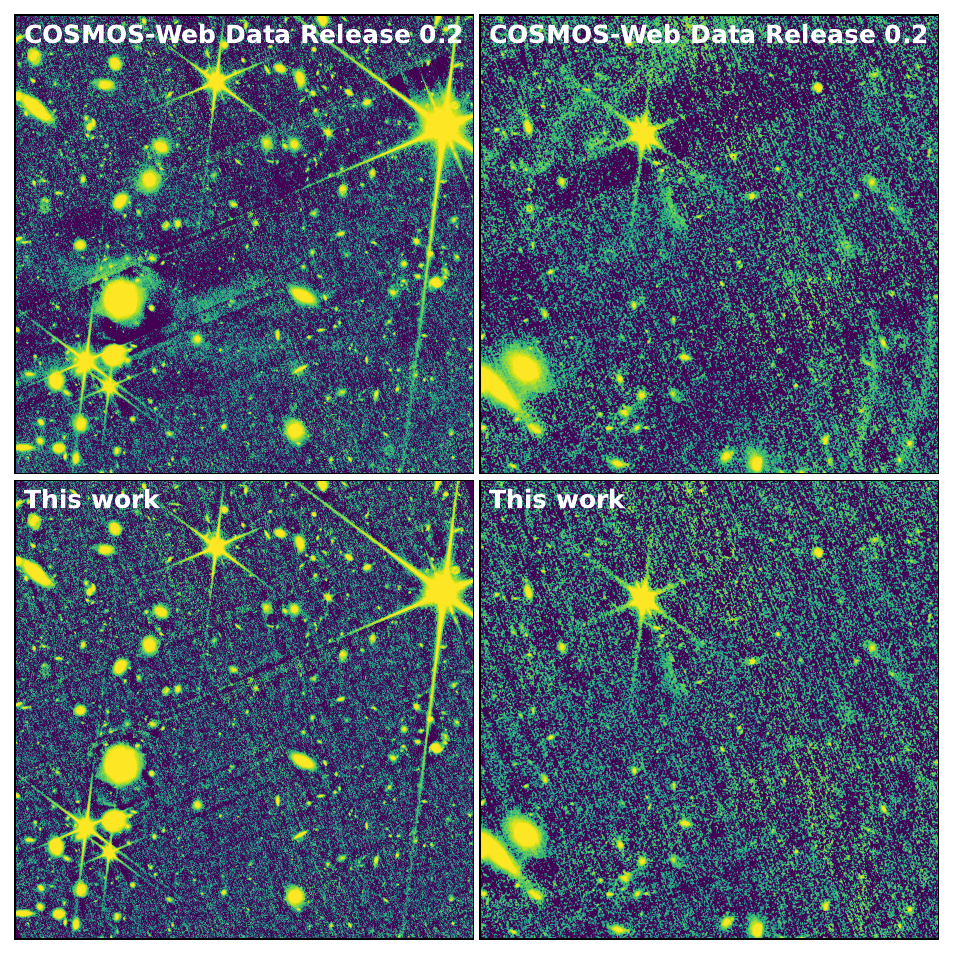}
\caption{Comparison of an F150W mosaic at Observation 047 between the public data release 0.2 by the COSMOS-Web team (top row) and our reduction (bottom row). Mosaics have been smoothed using a Gaussian kernel with a FWHM of 3 pixels. The left column highlights differences in background subtraction and the right column illustrates the removal of artifact features, i.e., wisps (near the bright star at the top of the image) and claws (near the right edge of the image). }
\label{fig2}
\end{figure}

Figure~\ref{fig2} compares our NIRCam F150W mosaic at Observation 047 with those from the 30~mas version of the public data release 0.2 by the COSMOS-Web team\footnote{\url{https://exchg.calet.org/cosmosweb-public/DR0.2/}}. Our mosaics show overall improved background subtraction, with significant improvement of wisps and claws removal. Average $5\sigma$ point source depths for areas covered by 4 dither-exposures (calculated within 0\farcs15 radius apertures without application of aperture corrections) are 27.40, 27.66, 28.34, and 28.25 AB mag in F115W, F150W, F277W, and F444W, respectively. These values are generally consistent with those reported in \citet{COSMOS-Web}.

\section{PSF Construction and Characterization} \label{Sec3}
\subsection{PSF Construction}
Accurate modeling of the PSF is crucial to robustly measuring the structural properties of galaxies and AGNs. Recently, \cite{Zhuang&Shen2023arXiv} have demonstrated that NIRCam mosaics have significant spatial PSF variations. The maximum and RMS fractional variations of PSF FWHM decrease from $\sim20$\% and 5\% in the F070W filter to $\sim3$\% and 0.6\% in the F444W filter, with exact values depending on the adopted resample parameters and dither pattern. They recommend the use of \texttt{PSFEx} \citep{Bertin2011ASPC} to model the PSF as well as its spatial variation given its superior performance over other commonly used methods. Therefore, we follow their procedures and construct PSF models consistently for both JWST NIRCam and HST ACS mosaics. We run SExtractor \citep{SExtractor} to select high SNR (\texttt{SNR\_WIN>100}), non-blended (\texttt{FLAGS<2}), non-irregular (\texttt{ELONGATION<1.5}), point-like (\texttt{CLASS\_STAR>0.8}), and without bad pixels (\texttt{IMAFLAGS\_ISO=0}). We further exclude sources with X-ray detection by matching with the Chandra COSMOS Legacy catalog \citep{Marchesi+2016ApJ} and 2.5$\sigma$ outliers in half-light radius (\texttt{FLUX\_RADIUS}). We adopt the \texttt{PIXEL\_AUTO} basis type and an oversampling factor of 2 (PSF model pixel scale is half the input value) with PSF model size of 201 pixels for HST ACS F814W and NIRCam SW filters and 301 pixels for NIRCam LW filters. 

For each NIRCam mosaic, we construct three types of PSF models: (1) a \textit{global} PSF model using all of the point-like sources across the entire FoV of each dither-combined mosaic; (2) a \textit{local} PSF model accounting for the spatial variation of PSF within the mosaic; (3) a set of two PSF models (\textit{broad} and \textit{narrow}), each of which is constructed using the broader or narrower half of the point sources divided by their median FWHM. Given the small amount of point sources in individual mosaic with median numbers of $\sim$34--40 sources across four filters, only a first order polynomial (linear) spatial-dependent model can provide usable \textit{local} PSF model (no artifacts and adequate SNR).

For HST ACS F814W images, we only construct one PSF model for each target due to the adoption of HST ACS mosaic product. We select all point-like sources located within a radius of 1.5 arcmin of the target (Section~\ref{Sec4}) and construct a \textit{global} PSF model using PSFEx following the same procedures as for NIRCam mosaics. A radius of 1.5 arcmin\footnote{We increase this radius to 2 arcmin for CID-27 due to the lack of available point-like sources.} is chosen to roughly match the field-of-view of 202 $\times$ 202 arcsec$^2$ for HST ACS exposures to minimize PSF variation, with the caveat that the source may not lie at the center and thus the point-like sources nearby may come from different observations. As a result, a median of 13 point-like sources are available for PSF construction for each AGN target. 

\begin{figure}[t]
\centering
\includegraphics[width=0.5\textwidth]{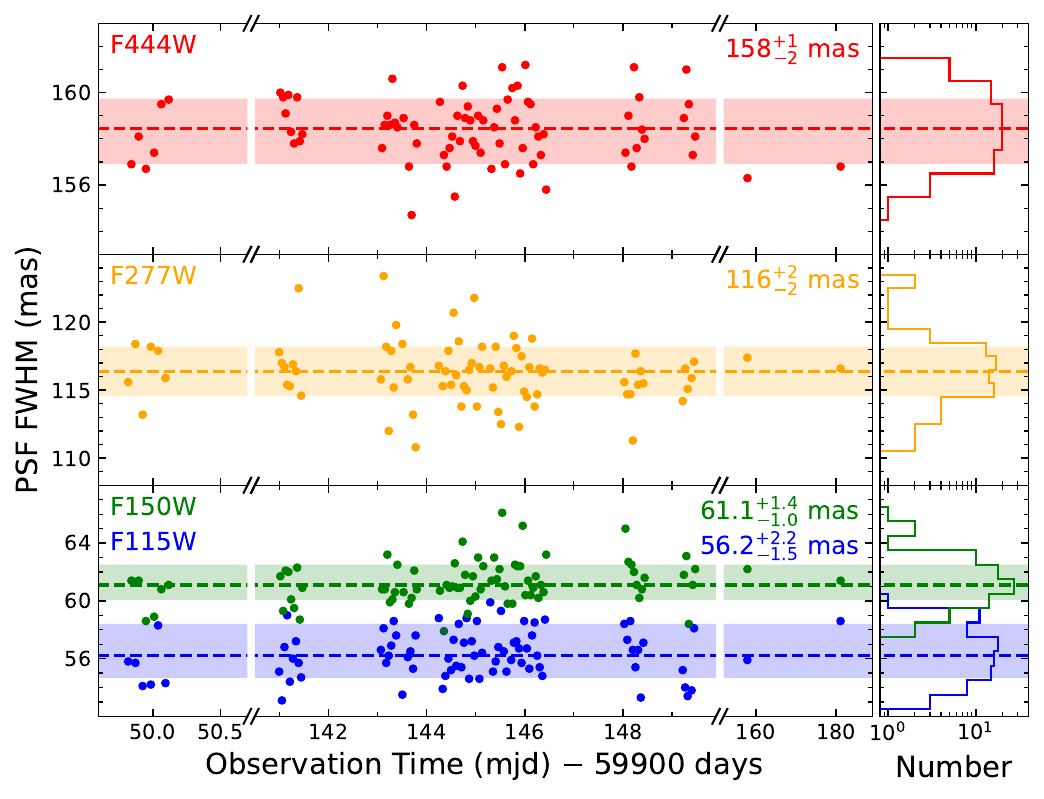}
\caption{PSF FWHM (\textit{global} model) versus observation time for four JWST NIRCam filters. Blue, green, orange, and red colors indicate F115W, F150W, F277W, and F444W, respectively. Horizontal dashed lines and shaded area represent median PSF FWHM and 16th to 84th percentiles, with the corresponding statistics shown at the upper-right corner. Right panels present histograms of PSF FWHM.}
\label{fig3}
\end{figure}

\subsection{PSF Properties}\label{Sec3.2}

Assuming that the \textit{global} PSF model represents typical PSF in each mosaic, Figure~\ref{fig3} shows the PSF FWHM of each observation as a function of observation time. FWHM of a PSF model is measured by fitting its core with a two-dimensional elliptical Gaussian function following \cite{Zhuang&Shen2023arXiv}. For all observations as a whole, the median PSF FWHMs of the F115W, F150W, F277W, and F444W filters are $56.2^{+2.2}_{-1.5}$, $61.1^{+1.4}_{-1.0}$, $116^{+2}_{-2}$, and $158^{+1}_{-2}$ mas, respectively, with super- and sub-script indicating the difference between 84th and 16th percentiles and the median (Table~\ref{table1}). These PSF FWHMs are much improved ($\sim$7--15\% for F115W and $\sim$6--13\% for F150W) for SW filters and slightly smaller for LW filters ($\sim$3--6\% for F277W and $\sim$1--2\% for F444W) compared to those constructed from point-like sources in mosaics with the same pixelscale but larger pixel ``shrunk'' fraction (\texttt{pixfrac=1}) \citep{Finkelstein+2023ApJ, Zhuang&Shen2023arXiv}. The temporal variation of PSF FWHM is dominated by short timescale fluctuations, with fractional root mean square (RMS) of $\sim2.9$\%, 2.4\%, 1.9\%, and 0.8\% for the F115W, F150W, F277W, and F444W filters, respectively (Table~\ref{table1}). The temporal variation in terms of FWHM is smaller compared to pixel-level flux fluctuations (on the order of $\sim$3--4\%) using calibrated individual exposures of two epochs presented in \cite{Nardiello+2022MNRAS}. 

We estimate the spatial variation across the field-of-view of an individual observation by measuring the RMS of FWHMs of \textit{local} PSF models at 100 positions randomly distributed across the field-of-view. The median 3$\sigma$-clipped fractional variations over all 80 visits in the F115W, F150W, F277W, and F444W filters are $\sim$2.8\%, 2.1\%, 2.7\%, and 1.0\%, respectively. These values are comparable with the temporal variation mentioned above, except for F277W, which is 0.8\% larger. 

As an independent check of the spatial variations estimated using \textit{local} PSF models, we directly measure the FWHM of point-like sources used for PSF model construction following the same method as for PSF models. With the caveat that the median values of FWHM may be slightly different due to different pixelscales for PSF model (0\farcs015) and mosaics (0\farcs03) and that the FWHM RMS may be affected by the spatial distribution of sources, we assume that the ratio between RMS and median of PSF FWHM remains the same. We find that the spatial variations of point-like sources in individual observations are higher than that estimated from randomly placed \textit{local} PSF models, with median 3$\sigma$-clipped fractional variations of $\sim$7.1\%, 5.0\%, 2.4\%, and 1.5\% among 80 visits (Table~\ref{table1}). In other words, the \textit{local} PSF models underestimate the true spatial variations of the PSF. 

To understand the differences between the two approaches, we take a closer look at the distribution of PSF FWHM in the entire field-of-view of each mosaic. We do not find a clear pattern with location in individual observations nor in a combined map constructed using all of the sources in 70 visits with the same JWST V3 position angle (107\degree). After binning the FWHM of sources in the combined map with bin sizes ranging from 50 to 500 pixels, we find a gradual reduction of the PSF FWHM RMS in the binned maps, with a reduction of $\sim$50\% for the bin size of 500 pixels. This suggests that the spatial variation in individual observations is dominated by random variations. This may be a result of the adopted dither pattern of COSMOS-Web (Figure~2 in \cite{COSMOS-Web}), which results in coverage with different number of exposures and detectors from different Modules. 

To summarize, we find that the median PSF FWHMs in our COSMOS-Web NIRCam mosaics are 56.2, 61.1, 116, and 158 mas for the F115W, F150W, F277W, and F444W filters, respectively. The PSFs of NIRCam imaging have significant temporal and spatial variations in terms of FWHM. The temporal variation of PSF FWHM decreases from $\sim$2.8\% (F115W) to 0.8\% (F444W) and is dominated by short timescale fluctuation. The spatial variation of PSF FWHM in individual observations are much larger compared to the temporal variation and dominated by random variation, with $\gtrsim5$\% for SW filters and $\sim2$\% for LW filters. The linear spatial-dependent model (i.e., the \textit{local} PSF model) is not accurate enough to model the random spatial PSF variations in individual observations and thus is not recommended for further analysis. Besides variations due to instrumental effects, the stochastic aliasing of the PSF in the F115W and F150W filters, the number of sources used for PSF construction, the dither pattern, and the accuracy of astrometry alignment for individual sources in different exposures may contribute to the observed temporal and spatial variations. We recommend the use of \textit{global}, \textit{broad}, and \textit{narrow} PSF models for COSMOS-Web NIRCam mosaics. The PSF models along with reduced mosaics of each observation in each filter will be made publicly available to the community at \url{https://ariel.astro.illinois.edu/cosmos_web/}. Users can adopt the one that best models the profile of the target of interest.

For HST ACS mosaics, the FWHM of the derived PSF models of our objects ranges from 94 to 104 mas, with a median of 98 mas and a RMS of 2 mas. The amount of fractional variation of HST ACS PSF FWHM ($\sim2\%$) is between that of F150W and F277W filters of JWST NIRCam. 

\begin{deluxetable}{ccccc}
\caption{Properties of NIRCam PSF \label{table1}}
\tabletypesize{\small}
\tablehead{
\colhead{Filter} & \colhead{F115W} & \colhead{F150W} & \colhead{F277W} & \colhead{F444W}
}
\startdata
PSF FWHM (mas) & 56.2 & 61.1 & 116 & 158 \\
Temporal variation & 2.8\% & 2.4\% & 1.9\% & 0.8\%\\
Spatial variation & 7.1\% & 5.0\% & 2.4\% & 1.5\%\\
\enddata
\tablecomments{PSF FWHMs and temporal variations are based on the \textit{global} PSF model of individual observations. Spatial variations are based on point-like sources in the mosaics.}
\end{deluxetable}

\begin{figure}[t]
\centering
\includegraphics[width=0.5\textwidth]{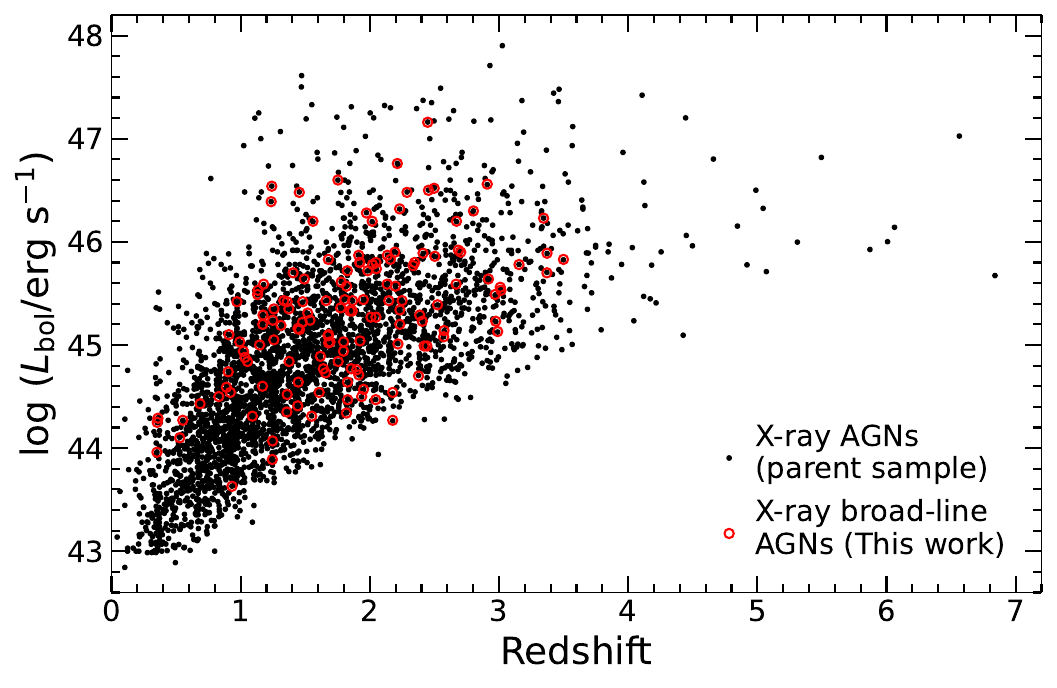}
\caption{Bolometric luminosity ($L_{\rm bol}$) versus redshift for the 143 X-ray broad-line AGNs (red circle) within the current COSMOS-Web footprint, and the parent X-ray AGN sample from \citet{Marchesi+2016ApJ} containing all AGN sources with either photometric or spectroscopic redshifts and $L_X \geq 10^{42}$ erg s$^{-1}$ in the entire COSMOS field.}
\label{fig4}
\end{figure}

\begin{figure*}[t]
\centering
\includegraphics[width=\textwidth]{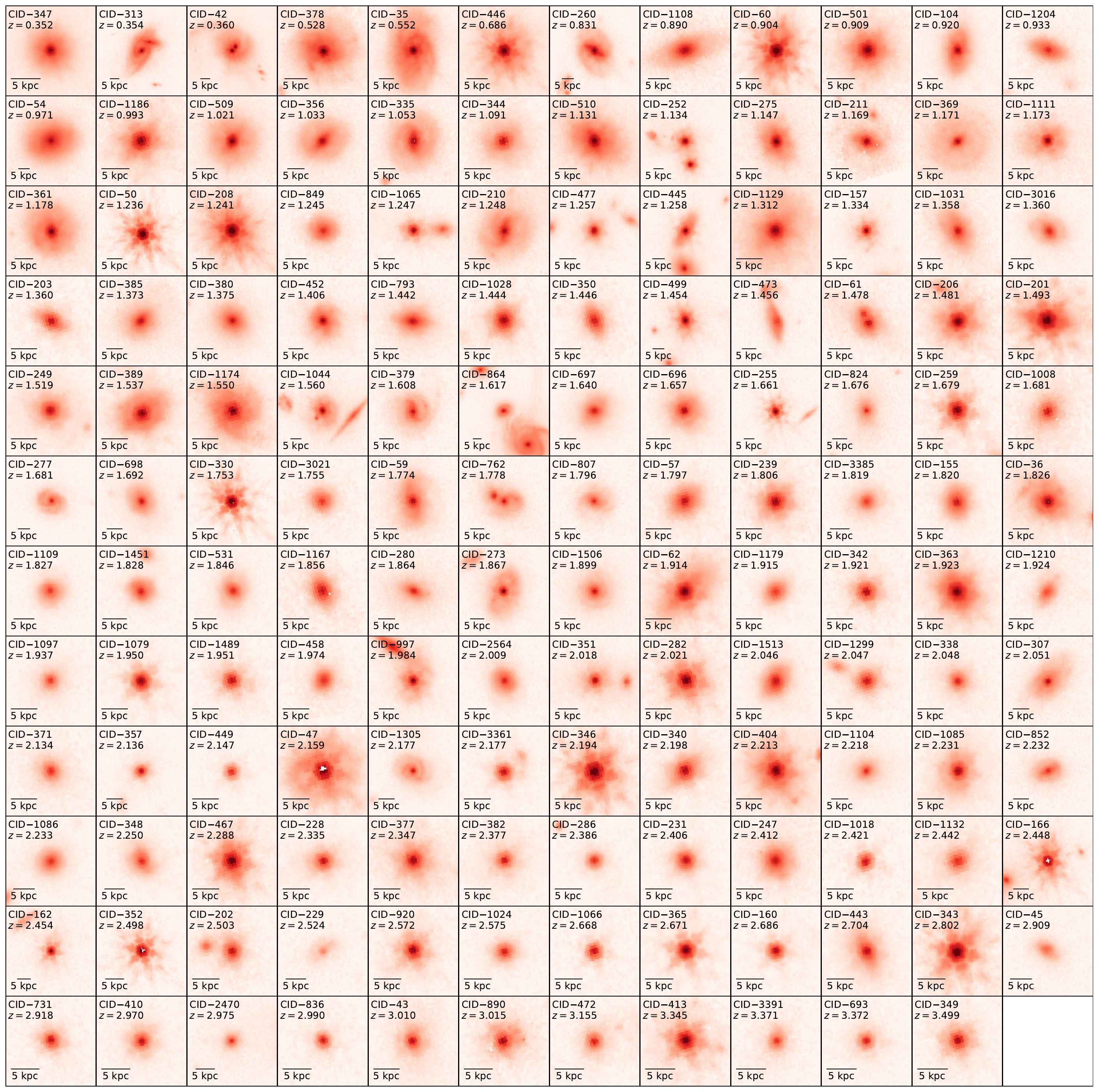}
\caption{A gallery of F277W images of our AGN sample in the order of ascending redshift. Object name and redshift are shown at the top and the scalebar of 5 kpc is shown at the lower-left corner. Note some nuclei have saturated central pixels, e.g., CID-47, CID-166, and CID-352.}
\label{fig5}
\end{figure*}

\begin{deluxetable*}{lcccccCCCCCc}
\tabletypesize{\footnotesize}
\caption{Properties of X-ray broad-line AGN sample \label{table2}}
\tablehead{
\colhead{Name} & \colhead{RA} & \colhead{DEC} & \colhead{$z$} & \colhead{log $L_{\rm X}$} & \colhead{log $L_{\rm bol}$} & \colhead{log $M_*$} & \colhead{$U-V$} & \colhead{$V-J$} & \colhead{$R_e$} & \colhead{$n$} & \colhead{Morph} \\
\nocolhead{Name} & \colhead{\degree} & \colhead{\degree} & \nocolhead{$z$} & \colhead{erg s$^{-1}$} & \colhead{erg s$^{-1}$} & \colhead{$M_{\odot}$} & \colhead{mag} & \colhead{mag} & \colhead{kpc} & \nocolhead{$n$} & \nocolhead{Morph} \\
\colhead{(1)} & \colhead{(2)} & \colhead{(3)} & \colhead{(4)} & \colhead{(5)} & \colhead{(6)} & \colhead{(7)} & \colhead{(8)} & \colhead{(9)} & \colhead{(10)} & \colhead{(11)} & \colhead{(12)}
}
\startdata
CID-1008 & 149.709109 & 2.089163 & 1.681 & 43.86 & 45.02 & 10.49 \pm 0.10 & 0.99 \pm 0.11 & 0.65 \pm 0.13 & 2.10 \pm 0.53 & 6.81 \pm 1.63 & \nodata \\
CID-1018 & 149.779096 & 1.928475 & 2.421 & 43.83 & 44.99 & 10.49 \pm 0.10 & 1.58 \pm 0.14 & 0.77 \pm 0.15 & 0.48 \pm 0.13 & 1.21 \pm 0.31 & \nodata \\
CID-1024 & 149.865785 & 2.027671 & 2.575 & 43.95 & 45.14 & 10.64 \pm 0.11 & 1.69 \pm 0.10 & 0.89 \pm 0.15 & 1.09 \pm 0.25 & 1.80 \pm 0.42 & \nodata \\
CID-1028 & 150.161767 & 1.877911 & 1.444 & 43.97 & 45.16 & 10.76 \pm 0.11 & 2.00 \pm 0.10 & 1.25 \pm 0.12 & 1.08 \pm 0.25 & 1.34 \pm 0.31 & \nodata \\
CID-1031 & 150.107862 & 1.861495 & 1.358 & 43.30 & 44.35 & 10.67 \pm 0.13 & 1.34 \pm 0.20 & 1.01 \pm 0.12 & 3.75 \pm 0.87 & 2.03 \pm 0.47 & S \\
CID-104  & 150.021048 & 2.420369 & 0.920 & 43.46 & 44.54 & 11.09 \pm 0.10 & 2.42 \pm 0.15 & 1.78 \pm 0.14 & 6.29 \pm 1.48 & 2.00 \pm 0.46 & S \\
CID-1044 & 150.245173 & 1.900061 & 1.560 & 44.67 & 46.20 & 11.31 \pm 0.11 & 1.75 \pm 0.17 & 1.10 \pm 0.15 & 5.12 \pm 1.18 & 4.49 \pm 1.04 & SI? \\
CID-1065 & 149.765899 & 2.054624 & 1.247 & 43.05 & 44.07 & 10.20 \pm 0.12 & 1.00 \pm 0.15 & 0.77 \pm 0.17 & 1.27 \pm 0.29 & 2.81 \pm 0.66 & I? \\
CID-1066 & 149.794429 & 2.073077 & 2.668 & 44.28 & 45.59 & 10.55 \pm 0.22 & 1.29 \pm 0.39 & 0.87 \pm 0.25 & 0.29 \pm 0.11 & 0.95 \pm 0.35 & \nodata \\
CID-1079 & 149.959995 & 2.032414 & 1.950 & 44.18 & 45.44 & 10.73 \pm 0.13 & 0.75 \pm 0.17 & 0.49 \pm 0.21 & 0.15 \pm 0.04 & 2.47 \pm 0.58 & \nodata
\enddata
\tablecomments{Col. (1): Object name. Cols. (2-3): Right ascension and declination in the F444W filter. Col. (3): Redshift. Col. (5): Rest-frame 2--10 keV absorption-corrected X-ray luminosity. Col. (6): Bolometric luminosity converted from \lx. Col. (7): Stellar mass. Cols. (8-9): Rest-frame $U-V$ and $V-J$ colors derived from SED fitting. Col. (10): Host effective radius at rest-frame 5000\AA. Col. (11): Host \sersic\ index at rest-frame 1\micron. Col. (12): Visual morphology. B: bar; I: interacting/tidal features with ``?'' indicating tentative evidence; S: spiral arms; \nodata: no clear feature. (Table~\ref{table2} is published in its entirety in the machine-readable format.)}
\end{deluxetable*}

\section{The AGN Sample}\label{Sec4}
The Chandra COSMOS-Legacy survey \citep{Civano+2016ApJ} is a 4.6 Ms project that covers 2.15 deg$^2$ of the COSMOS field with a flux limit of $\sim2\times 10^{-16}$ erg s$^{-1}$ in the 0.5--2 keV band. The identification of optical and NIR counterparts of 4016 X-ray sources, the collection and measurements of their redshifts, X-ray spectral properties (e.g., flux, hardness ratio), and AGN type classification are presented in \cite{Marchesi+2016ApJ}. In this work, we use the 632 spectroscopically-confirmed broad-line AGNs (\texttt{Clsp} $=1$) with reliable redshifts (\texttt{q\_zspec} $\geq1.5$) as our parent sample. Among them, 145 fall in the current footprint of COSMOS-Web NIRCam imaging. We exclude two sources (CID-58 and CID-161) that are contaminated by the spikes of nearby bright stars. Redshifts of CID-104 and CID-1305 are incorrect and are manually corrected to $z=0.920$ and $z=2.177$ by examining their spectra from \cite{Hasinger+2018ApJ}. CID-1305 is also included in \citet{Suh+2020ApJ} with the correct redshift as ours. The optical counterpart of CID-1132 is incorrect and is corrected to the source located $\sim$1\farcs5 to the west, which is much closer to the Chandra X-ray coordinates ($\sim$0\farcs3). Due to small overlapping regions among adjacent observations, 9 objects have the majority of their emission covered by two observations, offering independent information for uncertainty estimates.

We calculate the rest-frame 2-10 keV intrinsic (absorption-corrected) luminosity (\lx) for our sample assuming a power-law spectrum with a photon index $\Gamma = 1.8$ as:
\begin{equation}
    \lx = 4\pi D_L (1+z)^{\Gamma-2} \frac{10^{2-\Gamma} - 2^{2-\Gamma}}{E_2^{2-\Gamma} - E_1^{2-\Gamma}}  C_{E_1,E_2} f_{E_1,E_2},
\end{equation}
where $f_{E_1,E_2}$ is the observed flux between energies $E_1$ and $E_2$ in keV, $D_L$ is the luminosity distance, and $C_{E_1,E_2}$ is the absorption correction factor given in the soft (0.5--2.0~keV), hard (2.0--10.0~keV), and full (0.5--10.0~keV) bands in \cite{Marchesi+2016ApJ}. To minimize the effect of absorption, we adopt a priority order of hard $>$ full $>$ soft band to derive \lx~when sources are detected in more than one bands. Note that although our sample consists of optical type 1 AGNs, their X-ray emission could still be obscured, and a correction factor estimated from a simple hardness ratio value in \cite{Marchesi+2016ApJ} may not be appropriate \citep{Li2019}. Therefore, we also collect \lx~and \nh~from \cite{Marchesi2016_spectra} and \cite{Lanzuisi2018}, in which detailed X-ray spectral fitting is performed using an absorbed power-law model and the physical MYTorus model \citep{mytorus}, respectively, for a subset of the \cite{Marchesi+2016ApJ} sample. Our final \lx\ is based on the priority order of \cite{Lanzuisi2018} $>$ \cite{Marchesi2016_spectra} $>$ \cite{Marchesi+2016ApJ} (1, 113, and 29 sources, respectively). The \lx~of one source, CID-45, comes from the Compton-thick AGN sample in \cite{Lanzuisi2018}. The heavy obscuration  inferred from its X-ray spectrum is consistent with very low AGN contribution ($\lesssim2$\%) from our image decomposition (Section \ref{Sec5.1}). 
Moreover, we do not find concrete evidence for the presence of broad lines in CID-45 after visual inspection of its optical spectrum from \cite{Hasinger+2018ApJ}. Even so, we keep this object in our sample as the properties of its host galaxy remain robust. For CID-104 and CID-1305, their \lx\ values are derived from full and hard band flux, respectively, using the updated redshifts without absorption correction, since their correction factors are derived using the wrong redshifts in \citet{Marchesi+2016ApJ}. The presence of a bright nucleus in all five filters of the two objects is consistent with little absorption. 

We convert rest-frame 2--10 keV X-ray luminosity to bolometric luminosity ($L_{\rm bol}$) adopting luminosity-dependent bolometric correction from \citet{Lusso+2012MNRAS}, as parameterized by \citet{Yang+2018MNRAS}. The final sample consists of 143 broad-line AGNs spanning redshift range of $z=0.35-3.5$ (median $z=1.8$) and log $L_{\rm bol}$ range of 43.6--47.2 (median 45.3) erg s$^{-1}$. Figure~\ref{fig4} compares the distribution of redshift and $L_{\rm bol}$ of our X-ray broad-line AGN sample with the parent X-ray AGN sample from \citet{Marchesi+2016ApJ} containing all the sources (including those with photometric redshifts and type 2 AGNs) with $L_X \geq10^{42}$ erg s$^{-1}$. A gallery of F277W images of our sample is shown in Figure~\ref{fig5}. The properties of our sample are summarized in Table~\ref{table2}.

\begin{figure*}[t]
\centering
\includegraphics[width=0.9\textwidth]{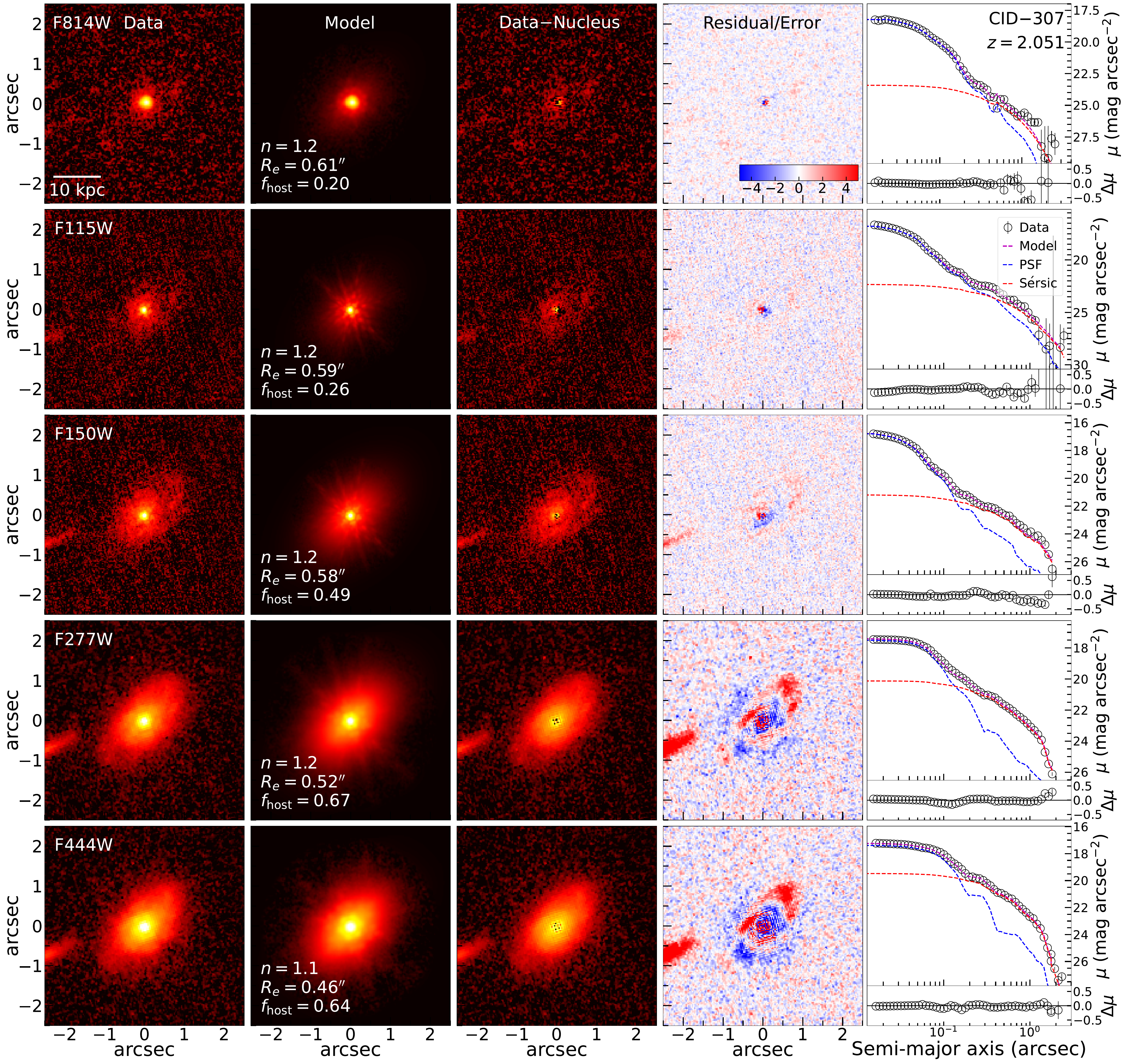}
\caption{Multiwavelength simultaneous AGN+host image decomposition to five-band ACS+NIRCam images of the X-ray broad-line AGN CID-307 at $z=2.051$ using \texttt{GALFITM}. Images of data, model (AGN+host), data$-$nucleus, and residual (data$-$model)/error are shown from left to right. The best-fit parameters for the host galaxy and host-to-total flux fraction $f_{\rm host}$ are shown at the lower-left corner in the model panel. The right-most column shows the surface brightness ($\mu$) profiles of data and models along the major axis, with the bottom panel showing the difference between data and model ($\Delta \mu=\mu_{\rm data}-\mu_{\rm model}$).}
\label{fig6}
\end{figure*}

\begin{figure}[t]
\centering
\includegraphics[width=0.5\textwidth]{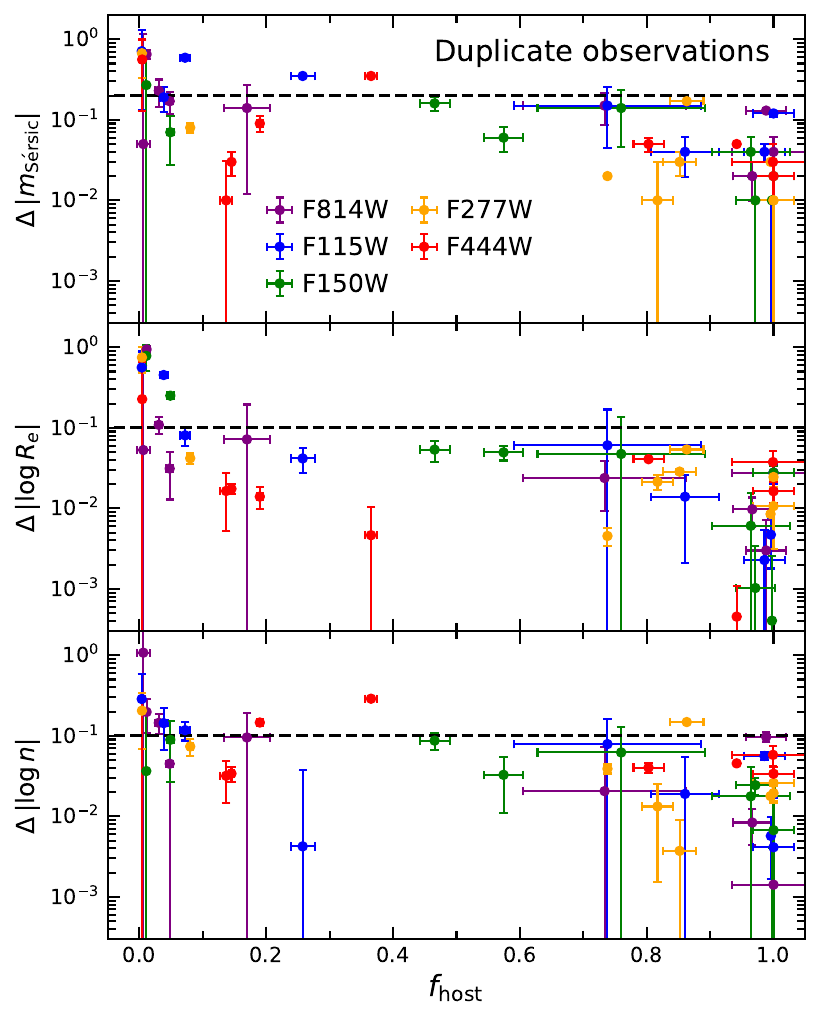}
\caption{Absolute difference of host parameters (\msersic: upper row, $R_e$: middle row, and $n$: bottom row) in duplicate observations of same objects versus host-to-total fraction ($f_{\rm host}$). Purple, blue, green, orange, and red dots and errorbars represent measurements and their uncertainties reported by \texttt{GALFITM} in the F814W, F115W, F150W, F277W, and F444W filters, respectively. Black dashed horizontal lines indicate the typical uncertainties added in quadrature to the nominal error reported by \texttt{GALFITM}. See detailed descriptions in the text. }
\label{fig7}
\end{figure}

\section{Properties of AGN Host Galaxies} \label{Sec5}

\subsection{Multiwavelength Simultaneous AGN-Host Image Decomposition}\label{Sec5.1}

We perform multi-wavelength simultaneous AGN-host image decomposition using \texttt{GALFITM} \citep{Haussler+2013MNRAS}, which is a multi-wavelength version of \texttt{GALFIT} \citep{Peng+2002AJ, Peng+2010AJ}. \texttt{GALFITM} takes into account the wavelength-dependent galaxy structure that is mainly due to different spatial distributions of stellar populations, metallicity, and dust attenuation. \texttt{GALFITM} is widely adopted to study the structure of galaxies of all redshifts \citep[e.g.,][]{Haussler+2022A&A, Kartaltepe+2023ApJ, Gillman+2023A&A} and to decompose the host galaxies of AGNs in the local universe and at cosmic noon \citep{Zhuang&Ho2022ApJ, Zhuang&Shen2023arXiv, Zhuang&Ho2023}

As our objects span a wide range of redshift ($z\approx 0.35-3.5$, corresponding to angular scales of 4.9--8.5 kpc arcsec$^{-1}$), we only adopt a PSF model for the AGN and a single \sersic\ model for its host galaxy for consistence in this work. We will study their substructures (e.g., bulges and bars) in a follow-up paper. We use a cutout size of the larger value between 20 times \texttt{FLUX\_RADIUS} in the F444W filter and 101 $\times$ 101 pixels ($\sim$3\arcsec). Blended companions are fitted simultaneously using either a \sersic\ model if \texttt{CLASS\_STAR<0.9} in all five filters or a PSF model. All other non-blended sources in the cutout are masked. The centers of the AGN and the host are tied together and allowed to vary independently in different filters to account for residual astrometry mismatches ($\lesssim$0.3 pixel among NIRCam images and $\sim$1.3 pixel between HST ACS F814W and NIRCam F277W). We also perform decomposition with AGN and host centers free to vary. We find consistent host parameters from decomposition regardless if we force the AGN and the host models to have the same center. We adopt constant ellipticity and position angle parameters across wavelength, leave magnitudes of the \sersic\ component (\msersic) and the AGN component ($m_{\rm AGN}$) free to vary, and allow the \sersic\ index $n$ and effective radius $R_e$ to vary quadratically with wavelength following \citet{Haussler+2013MNRAS} and \citet{Zhuang&Shen2023arXiv}\footnote{We restrict $R_e$ and $n$ to be constant across five filters in CID-166 due to unphysically large model host emission in the F814W and F115W filters.}. $n$ is restricted between 0.3 and 7 and $R_e$ between 0.5 and 100 pixels, corresponding to 0.07--25 kpc in the redshift range of our sample. 

We use measurements from SExtractor as initial guesses in \texttt{GALFITM}, with $m_{\rm AGN}$ and \msersic\ equal to 0.75 mag + total magnitude (assuming equal contribution from AGN and the host galaxy), $R_e$ equal to the half-light radius in the F277W filter, and $n$ equal to 1. For the sigma image, we use the error image from the \texttt{resample} step for JWST NIRCam images. As only a weight image (inverse variance of background) is available for ACS F814W mosaics, we construct sigma images by adding Poisson noise to the background noise from the weight image in quadrature, which preserves patterns from data reduction (such as cosmic ray flagging, flat fielding, and resampling). We note that although complicated substructures (e.g., bulge, bar, and spiral arms) are present in our sample (see Figure~\ref{fig12}), single \sersic\ model can recover the total galaxy emission as robust as more detailed models (e.g., bulge+disk) as suggested in studies of nearby galaxies \citep[e.g.,][]{Casura+2022MNRAS, Haussler+2022A&A}.

As we have three PSF models for each NIRCam filter (\textit{global}, \textit{broad}, and \textit{narrow}), we choose the best one for each filter by performing AGN-host decomposition in that filter. We adopt the \textit{global} PSF model by default unless improvement of reduced chi-squared ($\chi^2_{\nu}$) of the fit from \textit{broad} or \textit{narrow} PSF model is larger than 10\%. As a result, \textit{global} PSF models perform equally good or the best in all four filters in 134 sets of images. \textit{Broad} and \textit{narrow} PSF models perform significantly better in at least one filter in 11 and 7 sets of images, respectively. 

An example of our multiwavelength simultaneous AGN-host image decomposition is shown in Figure~\ref{fig6}. For all objects except CID-352, host galaxy can be clearly seen in the data$-$nucleus panel or surface brightness profiles in LW filters. We exclude CID-352 from the following analysis. Making use of results from SED fitting to AGN contamination-subtracted host fluxes (Section~\ref{Sec5.3}), we estimate the host-to-total fraction ($f_{\rm host}$) at rest-frame 5000\AA\ and 1 \micron. The 16th, 50th, and 84th percentiles of $f_{\rm host}$ are $\sim26$\%, 58\%, and 91\% at rest-frame 5000\AA\ and $\sim42$\%, 70\%, and 91\% at rest-frame 1 \micron.

\subsection{Uncertainty in AGN Host Parameters}

As discussed in previous works, the nominal error reported by the fitting code always underestimates the true uncertainty \citep[e.g.,][]{Haussler+2007ApJS, van_der_Wel2014ApJ}. Moreover, PSF mismatch, which is unavoidable in practice, could also lead to systematic biases to the recoverred host galaxy parameters \citep{Zhuang&Shen2023arXiv}. In this paper, we derive more reliable uncertainties for host parameters using duplicate observations of 9 AGNs in our sample, assuming their properties are representative among the full sample of 143. These 9 objects span a wide range of $f_{\rm host}$ from $\sim0$ to 1, \msersic\ from 18.5 to 24.3 mag, $R_e$ from 0\farcs02 to 1\farcs1, and $n$ from 0.3 to 7, covering the parameter space of the majority of our full sample. Figure~\ref{fig7} presents the absolute differences between two duplicate observations of host parameters (\msersic, $R_e$, and $n$) versus $f_{\rm host}$ in five filters. We find that host parameters derived from duplicate observations are in good agreement with each other across most of the $f_{\rm host}$ range except for $f_{\rm host}\lesssim5$\%, which is mainly driven by strong AGN contamination. The difference of host parameters between two independent observations include the effect of imperfect astrometry, pixel spatial sampling and PSF modeling, and thus provides a better estimate of the true uncertainty. Therefore, with the assumption that the results from these objects are representative of the parent sample, we adopt a fiducial systematic uncertainty of 0.2 mag for \msersic, 0.1 dex for $R_e$ and $n$ in all five filters to be added in quadrature to the uncertainty reported by \texttt{GALFITM}. Our adopted uncertainty of 0.2 mag for the host magnitude is generally consistent with that derived from mock AGN images by comparing the input and fitted parameters with similar $f_{\rm host}$ and host magnitude as our sample \citep{Zhuang&Shen2023arXiv}. We note that assuming a fixed uncertainty for the entire sample may lead to overestimated or underestimated error for individual objects. 

\subsection{Stellar Masses of AGN Host Galaxies} \label{Sec5.3}

We correct foreground Galactic extinction using the extinction curve from \citet{CCM1989} with $R_V=3.1$ and the dust map from \citet{Schlegel+1998ApJ}. We then adopt the SED fitting code \texttt{CIGALE v2022.1} \citep{Boquien+2019A&A, Yang+2022ApJ} to estimate the stellar masses of AGN host galaxies using fluxes after removing AGN contamination.

For the models used in \texttt{CIGALE}, we adopt a single stellar population model from \citet{Bruzual&Charlot2003MNRAS} with 0.2$\times$, 0.4$\times$, and $1\times$solar metallicity, a \citet{Chabrier2003PASP} IMF, and a ``delayed'' star formation history. We also use a \texttt{nebular} emission line model with ionization parameter $U=10^{-2}$, and an attenuation model from \texttt{dustatt\_modified\_starburst} with extinction curve adapted from \citet{Calzetti+2000ApJ}. We adopt a wide range of stellar ages: 1.0--10.0 Gyr in steps of 0.1 Gyr; an e-folding time of the main stellar population (Gyr): 0.001, 0.05, 0.1, 0.25, 0.5, 0.75, 1.0, 1.25, 1.5, 2.0, 2.5, 3.0, 3.5, 4.0, 4.5, 5.0, 6.0, 7.0, 8.0, 9.0, 10, 11, 12, 13, 14, 15, 16, 17, 18, 19, 20; and color excess of the nebular lines [$E(B-V)$ (mag)]: 0, 0.001, 0.005, 0.01, 0.03, 0.05, 0.1, 0.2, 0.3, 0.4, 0.5, 0.6, 0.7. All other parameters are fixed with their default values. 

To investigate the reliability of stellar mass estimates from SED fitting, we compare the stellar mass estimates from the best-fit template ($M_{*\\, {\rm best}}$) and those from the probability density function distribution of the likelihood of all the templates ($M_{*\\, {\rm bayes}}$). All objects have $|\log \frac{M_{*\\, {\rm best}}}{M_{*\\, {\rm bayes}}}|<0.3$ dex and $|{M_{*\\, {\rm best}}}-{M_{*\\, {\rm bayes}}}|\lesssim2\times$ uncertainty of $M_{*\\, {\rm bayes}}$, suggesting that stellar masses are self-consistent. All but six objects have $\chi^2_{\nu}\leq3$, indicating that our decomposition is reliable and robust for the vast majority of the sample. These six objects have excess host fluxes in the F814W or F115W filter likely due to an overestimation of host emission in objects with very low $f_{\rm host}$ \citep{Zhuang&Shen2023arXiv}. For CID-1065 with $\chi^2_{\nu}=2.5$, it has host flux excess compared to the best-fit SED template in the F444W filter likely due to underestimated AGN torus emission due to a compact host structure. We then fit host SEDs of these objects after excluding the filter with excess flux. The resulting stellar masses are consistent ($\lesssim 1\times$uncertainty) with the original values. For 9 objects with duplicate observations, their stellar masses are all consistent with each other. Therefore, we adopt the mean of the two estimates as their final stellar masses. We note that we do not include uncertainties from model assumptions, such as stellar initial mass function, stellar population model and star formation history, which can introduce $\sim0.3$ dex uncertainty to our stellar mass estimates \citep{Conroy2013ARA&A}.

The stellar masses of our sample span over $10^{10.1-11.5} M_{\odot}$ with a median of $10^{10.8} M_{\odot}$. We find a significant positive correlation between $M_*$ and $L_{\rm bol}$ (Spearman correlation strength $\rho=0.3$, $p$-value$=3\times10^{-4}$), with no correlation between $M_*$ and redshift. This is consistent with observations that the host galaxies of AGNs have abundant molecular gas reservoir, such that more massive hosts have more abundant gas content, capable of fueling more luminous AGN activities \citep[e.g.,][]{Beelen+2004A&A, Bischetti+2021A&A}.

\subsection{Stellar Mass Comparison with Previous Works}

We compare our stellar masses with those reported in earlier works that are derived using two approaches: (1) multiwavelength AGN+host SED fitting ({i.e., total AGN+host fluxes without performing image decomposition}) implementing AGN and galaxy templates \citep{Zou+2019ApJ,Suh+2020ApJ}\footnote{Redshift of CID-1174 is incorrect in \cite{Suh+2020ApJ} and, as a result, is excluded}; (2) mass-to-light ratio from host color from AGN+host image decomposition to HST images from \cite{Ding+2020ApJ}. After crossmatch, we obtain 95 objects from \cite{Zou+2019ApJ}, 26 objects from \cite{Suh+2020ApJ}, and 5 objects from \cite{Ding+2020ApJ} that are also covered by our AGN sample. We have corrected their stellar masses to match our adopted cosmology.

We first compare our stellar masses with those from fits to AGN+host SEDs using AGN and stellar templates \citep{Zou+2019ApJ, Suh+2020ApJ}. We find that our stellar masses are systematically larger than those from \cite{Zou+2019ApJ} with a median difference of $0.16^{+0.35}_{-0.24}$ dex (upper- and lower-script indicating differences between 84th and 16th percentiles from the median) and systematically lower than those from \cite{Suh+2020ApJ} with a median difference of $-0.34^{+0.20}_{-0.30}$ dex. Four objects in \cite{Zou+2019ApJ} and one object in \cite{Suh+2020ApJ} have absolute stellar mass difference larger than 1 dex from our measurements. We have verified that host emission from these five objects are robustly detected in our image decomposition and fits to their host SEDs are reliable and physical. 

As demonstrated in \citet{Zhuang&Ho2023} using mock AGNs generated from real galaxy SEDs, fitting composite AGN+host SEDs without image decomposition first can easily lead to overestimated or underestimated stellar masses as a result of the degeneracy between AGN templates and stellar templates. The systematic overestimation of $M_*$ increases with greater AGN contribution and bluer host color, and vise versa for the underestimation of $M_*$. On the other hand, inconsistent fluxes measured from surveys with different resolutions, sensitivities and apertures can lead to large scatter. Different choices of parameters of AGN and stellar templates can also lead to systematic differences, as illustrated by the large discrepancy between \cite{Zou+2019ApJ} and \cite{Suh+2020ApJ}. Our method, which independently decomposes AGN and host using high-resolution imaging, does not suffer from extra complications from adopting different AGN templates. Therefore, our comparison suggests that stellar masses estimated from fitting combined AGN+host SEDs are less reliable, with complicated systematics and have a small chance of catastrophic failures (deviation from the truth by more than 1 dex). 

For the five $M_*\gtrsim10^{10.7}M_{\odot}$ AGNs overlapping with \cite{Ding+2020ApJ}, 3 objects imaged with HST F140W filter have $M_*$ differences of $\sim0.3$ dex (ours are higher) and the other 2 imaged with HST F125W filter have differences of $\sim0$ and $-0.1$ dex. Many factors can contribute to the differences. At $z\approx1.5$, the F140W filter only probes rest-frame $\sim5600$\AA, missing the peak emissions from older stellar populations that contribute the majority of the stellar mass in these massive systems. As only two HST filters are available, \cite{Ding+2020ApJ} estimate stellar masses using mass-to-light ratios from stellar populations with fixed age (1 Gyr for $z<1.44$ and 0.625 Gyr for $z>1.44$), adding extra uncertainty to the measurements. As a comparison, the stellar age from our SED fitting ranges from 1.2--1.6 Gyr. The three objects with $\sim0.3$ dex stellar mass differences have $z>1.44$, which indicates that the difference is partly due to their adopted fixed stellar population template. Moreover, the much improved spatial resolution and sensitivity of JWST NIRCam imaging compared to HST makes our decomposition more reliable. By comparing $f_{\rm host}$ in HST F140W with NIRCam F150W and that in HST F125W with NIRCam F115W, we find an overestimation of $\sim10\pm6$\% in their decomposition to HST images.

\section{Results and Discussion}\label{Sec6}

\begin{figure}[t]
\centering
\includegraphics[width=0.5\textwidth]{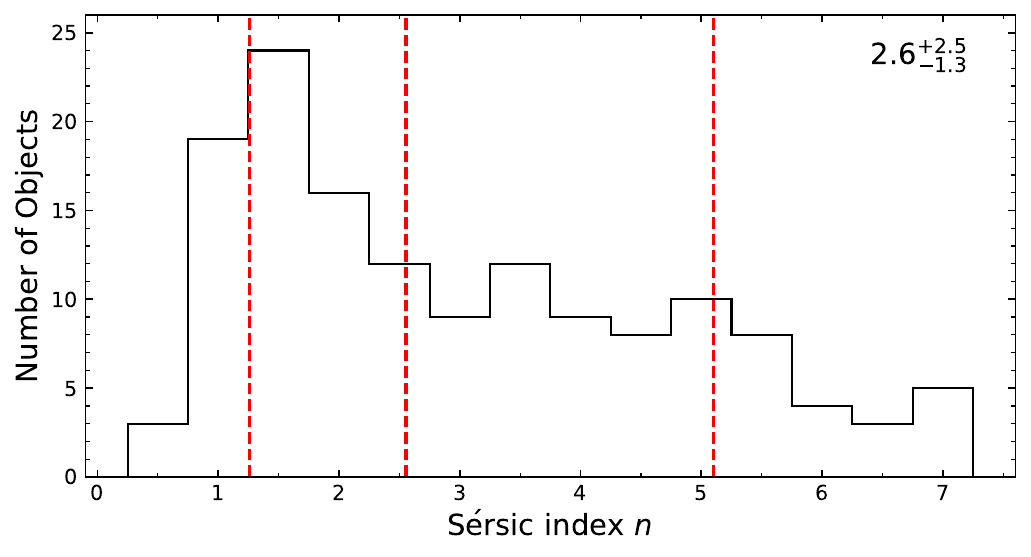}
\caption{Histogram of \sersic\ index $n$ at rest-frame 1 \micron. Red vertical dashed lines indicate 15th, 50th, and 84th percentiles of the sample with statistics shown at the upper-right corner.}
\label{fig8}
\end{figure}

\begin{figure}[t]
\centering
\includegraphics[width=0.5\textwidth]{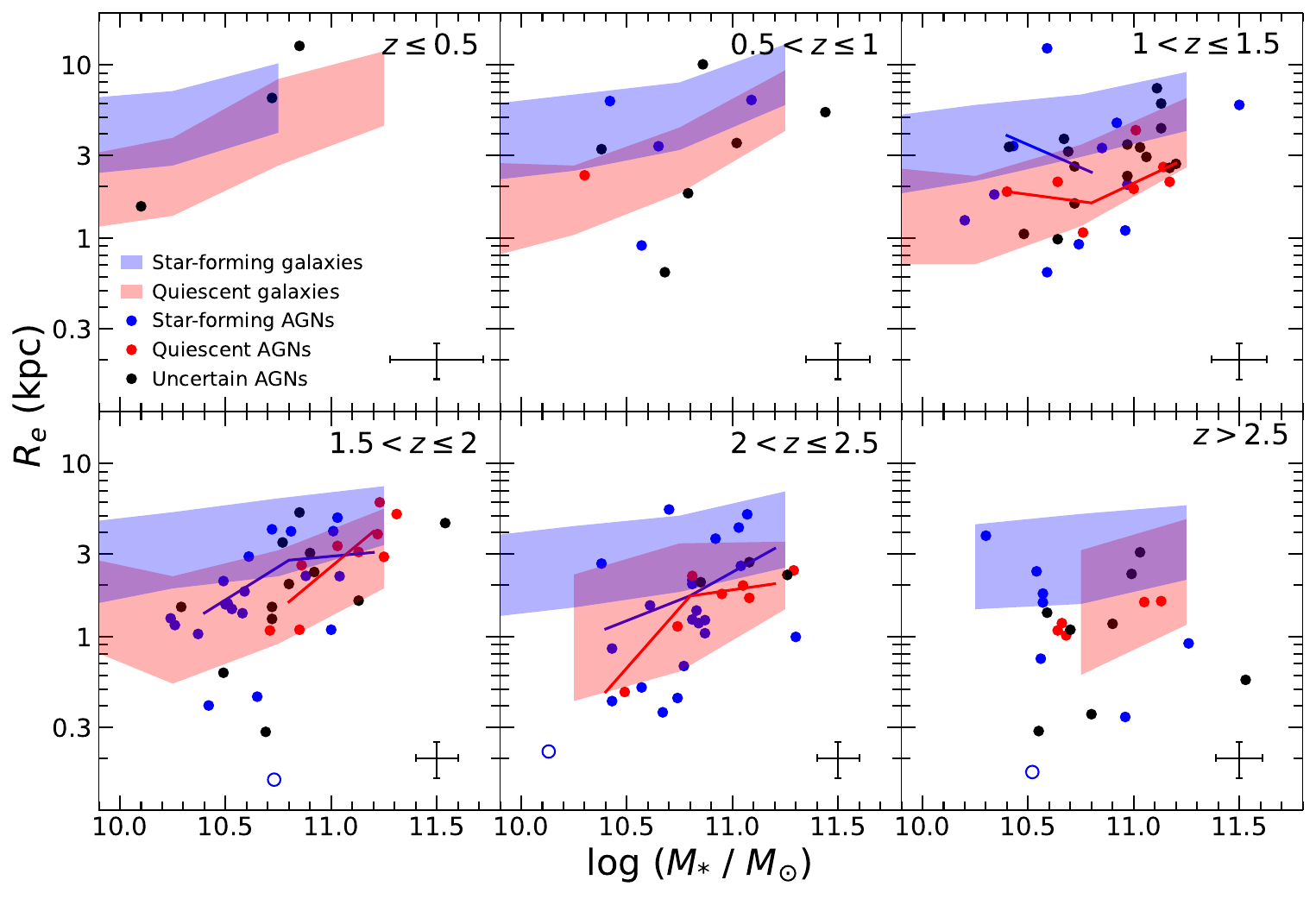}
\caption{Host stellar mass--size relation for X-ray-selected broad-line AGNs. Size is $R_e$ interpolated to rest-frame 5000\AA. Blue and red shaded areas represent the 16th--84th percentile distributions of star-forming and quiescent {non-AGN} galaxies from \cite{van_der_Wel2014ApJ}, respectively. Blue and red dots represent star-forming and quiescent AGN hosts classified using the UVJ diagram, while black dots represent AGNs without robust classification given the uncertainty of colors and proximity to the boundary. Open symbols indicate objects with $R_e<1$ pixel. Median sizes as a function of stellar mass are shown with solid curves for objects at $1<z\leq2.5$. Error bar indicates typical uncertainty.}
\label{fig9}
\end{figure}

\subsection{Color, Morphology and Structure of AGN Host Galaxies}

\subsubsection{AGN Hosts Are Mainly Star-forming Galaxies}

We classify the host galaxies of AGNs into star-forming and quiescent galaxies using the UVJ diagram from \cite{Williams+2009ApJ}. Rest-frame $U-V$ and $V-J$ colors are obtained from \texttt{CIGALE}. We find that the hosts of $86\pm4$ ($61\pm3$\%; considering the uncertainty of color) AGNs are classified as star-forming galaxies. A large fraction of AGN hosts are star-forming galaxies is consistent with previous studies of X-ray AGNs at similar redshifts \citep[e.g.,][]{Rosario+2013ApJ, Sun+2015ApJ, Mountrichas+2021A&A, Coleman+2022MNRAS} and those of luminous AGNs or quasars in the nearby Universe \citep[e.g.,][]{Jarvis+2020MNRAS, Xie+2021ApJ, Li+2021ApJ, Zhuang&Ho2022ApJ, Zhuang&Ho2023, Li+2023arXiv}. Our results support the mutual dependence of BH accretion and host star formation on molecular gas reservoir \citep[e.g.,][]{Shangguan+2020ApJ, Koss+2021ApJS, Zhuang+2021ApJ} and suggest no instantaneous negative AGN feedback \citep[e.g.,][]{Costa+2014MNRAS, Harrison+2018NatAs}.

\subsubsection{\sersic\ Index Distribution}

We estimate the \sersic\ index $n$ at rest-frame 1 \micron\ by fitting a second-order polynomial to wavelength and $n$ of five filters, following our assumption of wavelength-dependence in the decomposition (Section~\ref{Sec5.1}). Figure~\ref{fig8} shows the histogram of \sersic\ index $n$ at rest-frame 1 \micron. We find that for X-ray broad-line AGNs, $n$ peaks around $1-2$, with a relatively flat distribution toward larger $n$. If we classify objects with $n\leq2$ as late-type disk-dominated galaxies and $n>2$ as early-type bulge-dominated galaxies \citep{Zhuang&Ho2023}, 54 AGNs ($\sim38$\%) live in late-type disk-dominated galaxies. We do not find a clear dependence of $n$ on stellar mass, with $\sim40$\% and $37$\% of late-type galaxies with $M_*<10^{10.8} M_{\odot}$ and $M_* \geq 10^{10.8} M_{\odot}$, respectively. The fraction remains robust if we switch to $n$ at rest-frame 5000\AA. This fraction is in broad agreement with AGNs in the nearby Universe and at $1<z<3$ \citep[e.g.,][]{Fan+2014ApJ, Rosario+2015A&A, Ding+2020ApJ, Bennert+2021ApJ, Kim+2021ApJS, Li+2021ApJ, Zhuang&Ho2023, Li_Jennifer_2023arXiv}, suggesting no significant redshift evolution of \sersic\ index of AGN host galaxies. 

We note that while low $n$ values ($\sim1$) indicate that the galaxy is dominated by stellar disk, large $n$ values ($n\gtrsim2$) does not necessarily rule out the presence of a disk. For example, a combination of a high surface brightness, small-scale (pseudo-)bulge and a low surface brightness, large-scale disk can also mimic the surface brightness profile with a large $n$. In a follow-up paper, we will perform detailed bulge(+bar)+disk decomposition to the AGN host galaxies to study the properties of bulge and disk and compare with those in a non-AGN control sample.

\subsubsection{Stellar Mass--size Relation}

The relation of stellar mass of a galaxy and the size distribution of its mass, stellar mass-size ($M_*-R_e$), encodes information of the galaxy evolution and assembly history. To investigate the position of X-ray broad-line AGNs on the $M_*-R_e$ plane, we estimate $R_e$ at rest-frame 0.5 \micron, following the same procedures as $n$ mentioned above. $R_e$ estimated using this method is in excellent agreement ($-0.01^{+0.04}_{-0.08}$ dex) with that estimated using color gradient in \cite{van_der_Wel2014ApJ}. For the $M_*-R_e$ relation, we use more accurate host color classification. We consider the classification of an object accurate only if it is classified as the same type (star-forming or quiescent) in at least 75\% of the cases after perturbing its $U-V$ and $V-J$ colors using their uncertainties in 100 trials. As a result, we have 66 star-forming hosts, 29 quiescent hosts, and 47 hosts without reliable classification. 

Figure~\ref{fig9} shows the $M_*-R_e$ relation of X-ray broad-line AGNs in different redshift bins, compared to the non-AGN galaxies in \citet{van_der_Wel2014ApJ}. The classification of star-forming and quiescent galaxies in \citet{van_der_Wel2014ApJ} is also based on the UVJ diagram from \citet{Williams+2009ApJ}. We find that AGN hosts generally follow the $M_*-R_e$ relation for non-AGN galaxies in the same redshift range \citep{van_der_Wel2014ApJ}. However, star-forming AGN hosts lie slightly lower compared to their non-AGN counterparts at $1<z\leq 2.5$, where we have more than 10 star-forming AGN hosts. This result, albeit with small statistics, echoes previous studies at similar redshifts \citep[e.g.,][]{Barro+2014ApJ, Silverman+2019ApJ} and supports the idea that a significant fraction of AGNs tends to be hosted by compact star-forming galaxies \citep[e.g.,][]{Li+2021ApJ}, which are associated with concentrated molecular gas and dust distribution as suggested by recent submm observations of AGNs at cosmic noon \citep[e.g.,][]{Stacey+2021MNRAS, Jones+2023MNRAS}. 

\begin{figure}[t]
\centering
\includegraphics[width=0.5\textwidth]{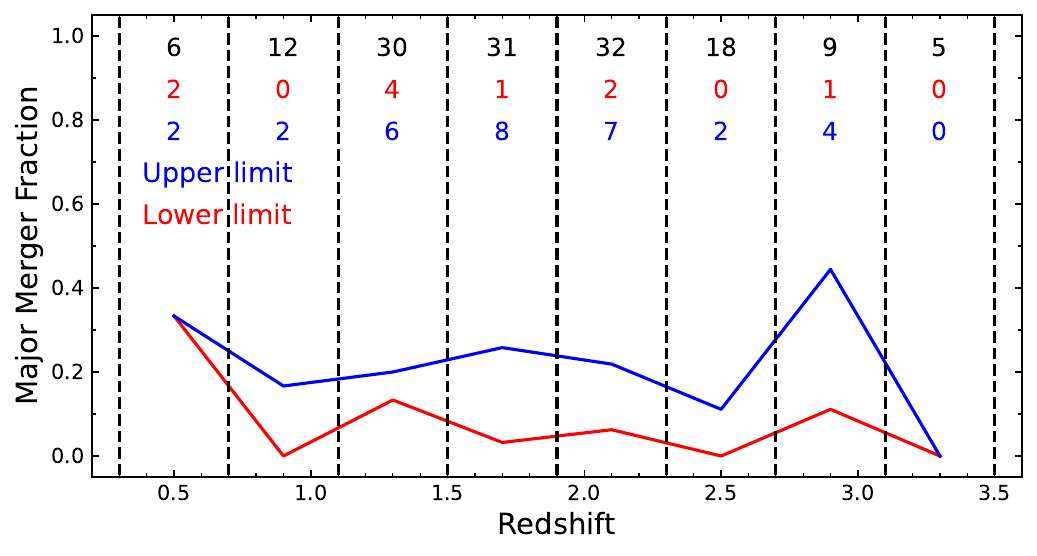}
\includegraphics[width=0.5\textwidth]{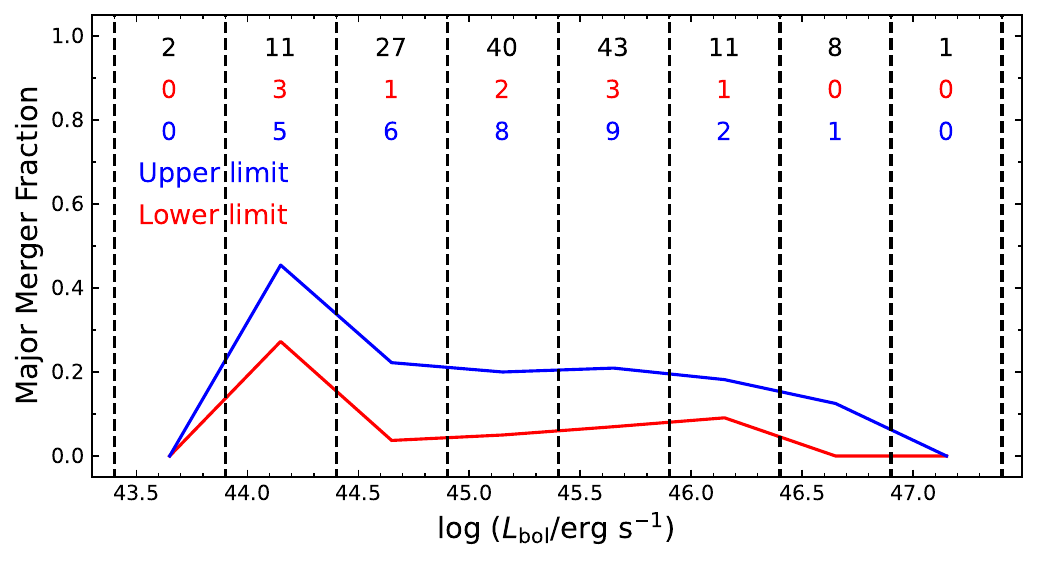}
\caption{Major merger fraction versus redshift (top panel) and AGN bolometric luminosity (bottom panel). Blue and red colors indicate upper and lower limit of major merger rate, respectively. Number of all objects (black), upper limit (blue) and lower limit (red) of major merger candidates in each bin (separated by black vertical dashed lines) are shown at the top.}
\label{fig10}
\end{figure}

\begin{figure*}[t]
\centering
\includegraphics[width=\textwidth]{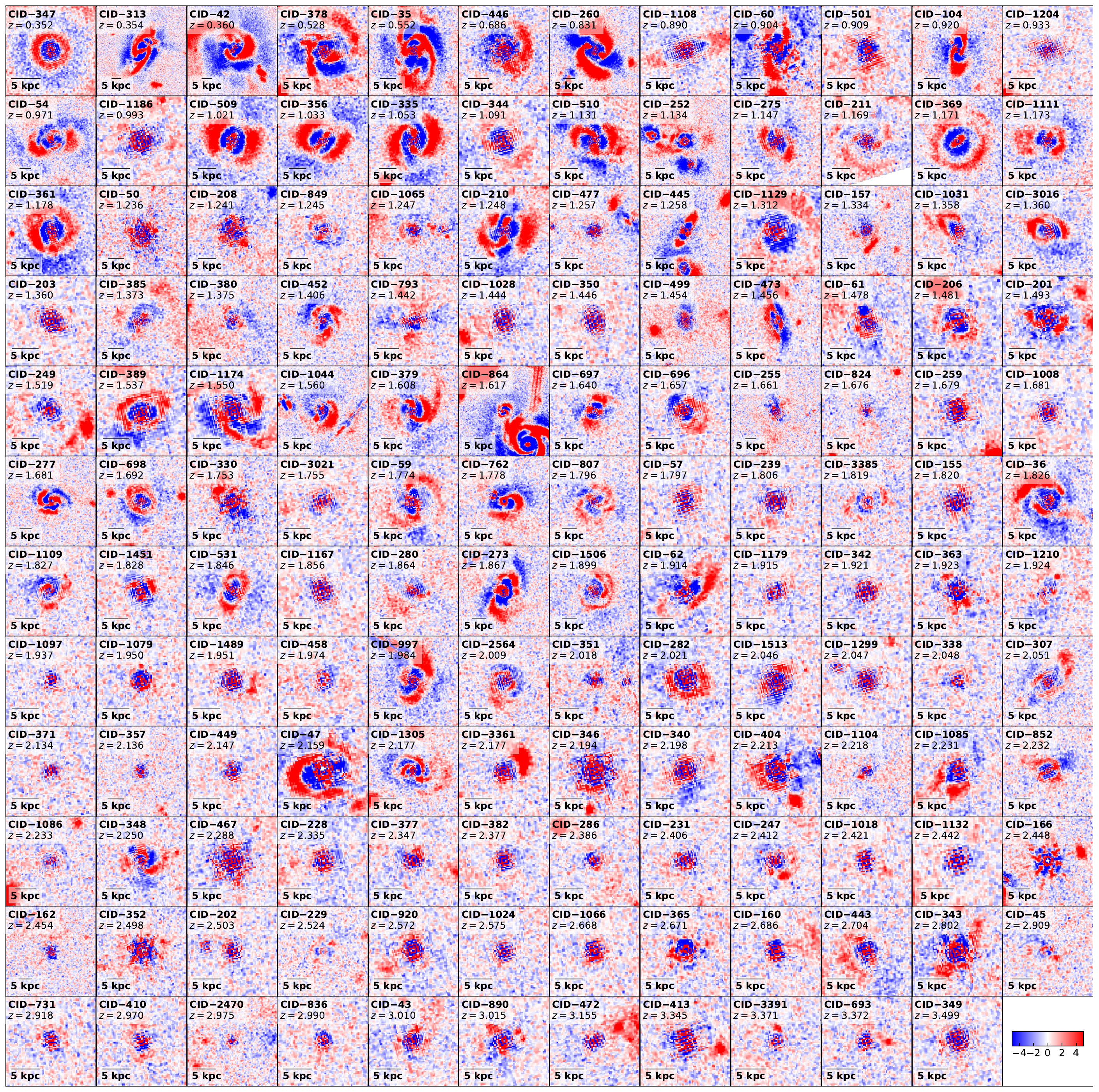}
\caption{A gallery of normalized residual images (residual/error) of our sample in the F277W filter after subtracting best-fit PSF and \sersic\ models in the order of ascending redshift. Object name and redshift are shown at the top and the scalebar of 5 kpc is shown at the lower-left corner. Colorbar is shown in the bottom-right panel. Non-axisymmetric features are ubiquitous in X-ray-selected broad-line AGN hosts.}
\label{fig11}
\end{figure*}

\subsection{The Ubiquitous Presence of Non-axisymmetric Features} \label{Sec6.2}

Gas-rich major mergers have long been considered as an important triggering mechanism of luminous AGNs, despite the longstanding debate of whether it is the primary triggering mechanism \citep[e.g.,][]{Alexander&Hickox2012NewAR, Treister+2012ApJ, Ellison+2019MNRAS}. Leveraging the sensitivity and rest-frame near-IR coverage of JWST NIRCam, we study the close environment of AGNs at cosmic noon. We select close major merger candidates by applying the following two criteria: (1) sources have clear evidences of interacting/tidal features in residual maps in any of five filters (25 candidates); (2) sources have companions with magnitude difference (compared to the magnitude of the AGN host) smaller than 1.5 mag (i.e., flux ratio of 1:4) in the F277W filter and within a projected distance of 20 kpc (16 candidates). With 5$\sigma$ detection limit of 27.7--28.3 mag in the F277W filter \citep{COSMOS-Web} and the faintest host of 23.3 mag, we can detect all potential companions if present. In total, 31 sources satisfy either criterion and 10 sources satisfy both criteria. We caution that candidates selected from the first criterion may contain interacting features caused by minor mergers, while those selected from the second criterion may not have physically associated companions since no reliable spectroscopic redshifts are available for the companions. Therefore, 31/143 ($\sim22$\%) should be considered as a strict upper limit of major merger fraction, while 10/143 (7\%) can be considered as a conservative lower limit. Figure~\ref{fig10} shows the major merger rate as a function of redshift and AGN strength. Given the limited sample size of this work, we do not find clear dependence of major merger rate on redshift and AGN strength. The overall major merger rate in AGNs is consistent with massive inactive galaxies at similar redshifts \citep{Bundy+2009ApJ, Duncan+2019ApJ, Ferreira+2020ApJ, Whitney+2021ApJ}. 

On the other hand, minor mergers and secular processes can also drive gas inflow and fuel BH accretion \citep[e.g.,][]{Kormendy&Kennicutt2004ARA&A, Kaviraj+2014MNRAS, Kim&Kim2014MNRAS, Shu2016ARA&A}. We visually check the normalized residual (data$-$PSF$-$\sersic) images (see Figure~\ref{fig11} for an example in the F277W filter) in all five filters and identify clear presence of stellar bar or spiral arms in 40 (28\%) objects. Besides the aforementioned strong non-axisymmetric features, nearly all the objects show various degrees of disturbed morphology or faint companion(s) (flux ratio $\ll 1:4$), which may be interpreted as evidence of minor mergers. The ubiquitous presence of non-axisymmetric features combined with a small major merger fraction (7--22\%) highlights the importance of gas fueling by secular processes and minor mergers, with caveat of small sample size investigated here. A quantitative comparison to a redshift-, stellar mass-, and galaxy structure-matched non-AGN sample and deep spectroscopic survey of companions are required to fully understand the role of different mechanisms in triggering AGN activities and if AGNs are special with respect to the general galaxy population.

\begin{figure}[t]
\centering
\includegraphics[width=0.5\textwidth]{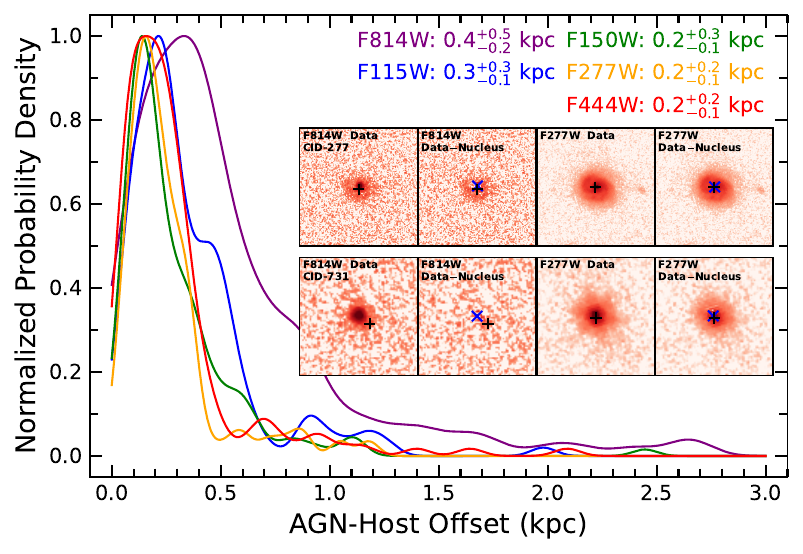}
\caption{Normalized probability density from Gaussian kernel-density estimate of AGN-host offset in the F814W (purple), F115W (blue), F150W (green), F277W (orange), and F444W (red) filters. Medians and the differences between 84th and 16th percentiles are shown at top-right corner. Inset figures show F814W data, F814W data$-$nucleus, F277W data, and F277W data$-$nucleus of CID-277 (top) and CID-731 (bottom) from left to right. Blue cross represents the location of AGN and black plus represents the location of host galaxy.}
\label{fig12}
\end{figure}

\subsection{AGN-Host Centroid Offset}

Off-nucleus AGNs may indicate the presence of close dual supermassive BHs in the inspiraling phase during galaxy merger or recoiled BHs from gravitational kickout of the binary coalesces \citep[e.g.,][]{Baker+2006ApJ, Barth+2008ApJ, Campanelli+2007PhRvL, Comerford&Greene2014ApJ}. We investigate the AGN host offset in our X-ray broad-line sample, utilizing the results from image decomposition results without tying the centers of the PSF and \sersic\ components. Here, the PSF-\sersic\ offset refers to that between the AGN and its associated stellar core, rather than between the AGN and a close companion in a double-core system, such as CID-42 \citep{Li+2023_CID42}. Figure~\ref{fig12} shows the normalized probability of AGN-host offset in five filters used in this work. We find that the median offsets in all five filters are all quite small ($\leq0.4$ kpc), suggesting good alignment of the AGN and the galaxy center. 

However, the median offset and the spread increase systematically toward shorter wavelength. This can be driven by physical reasons such as non-axisymmetric spatial distribution of extinction and stellar populations so that their presence may depend on wavelength. An example is shown at the upper inset of Figure~\ref{fig12}. Galaxies are more asymmetric and bright star-forming clumps are common at $z\gtrsim1$ \citep[e.g.,][]{Elmegreen+2007ApJ, Guo+2015ApJ}, which may be due to fragmentation of massive gas disks driven by gravitational instabilities \citep[e.g.,][]{Noguchi1999ApJ, Dekel+2009ApJ}. \cite{Wuyts+2012ApJ} find that off-center clumps can contribute up to $\sim20$\% to the integrated SFR, but only 7\% or less to the total mass in massive star-forming galaxies at $z\approx 1 - 2$, consistent with decreasing AGN host offset toward longer wavelength. On the other hand, AGN host offset can be entirely due to intrinsic faintness of the host associated with low $f_{\rm host}$ in rest-frame UV, as illustrated at the lower inset of Figure~\ref{fig12}. \citet{Zhuang&Shen2023arXiv} have demonstrated that artificial offsets are commonly found in hosts with low host surface brightness using mock AGNs. 

The observed residual AGN host offset in the F277W and F444W is small ($\sim0.2$ kpc, $<1$ pixel). These residual offsets could be due to the presence of nuclear substructure or PSF mismatch. The long tail of the distribution is due to a combination of faint nucleus and asymmetric light distribution from interactions (Section~\ref{Sec6.2}). Nonetheless, the high resolution of JWST NIRCam provides more stringent constraints on AGN host offset compared to previous results based on HST images and Gaia astrometry \citep[e.g.,][]{Li_Jennifer_2023arXiv, Shen+2019ApJ}. Therefore, our results suggest that there is no significant AGN host offset in the majority of X-ray broad-line AGNs.  Genuine detection of off-nucleus AGNs must require well-defined host centroid, and significant AGN-host centroid offset beyond systematic uncertainties. This topic will be addressed in a follow-up paper.

\begin{figure}[t]
\centering
\includegraphics[width=0.5\textwidth]{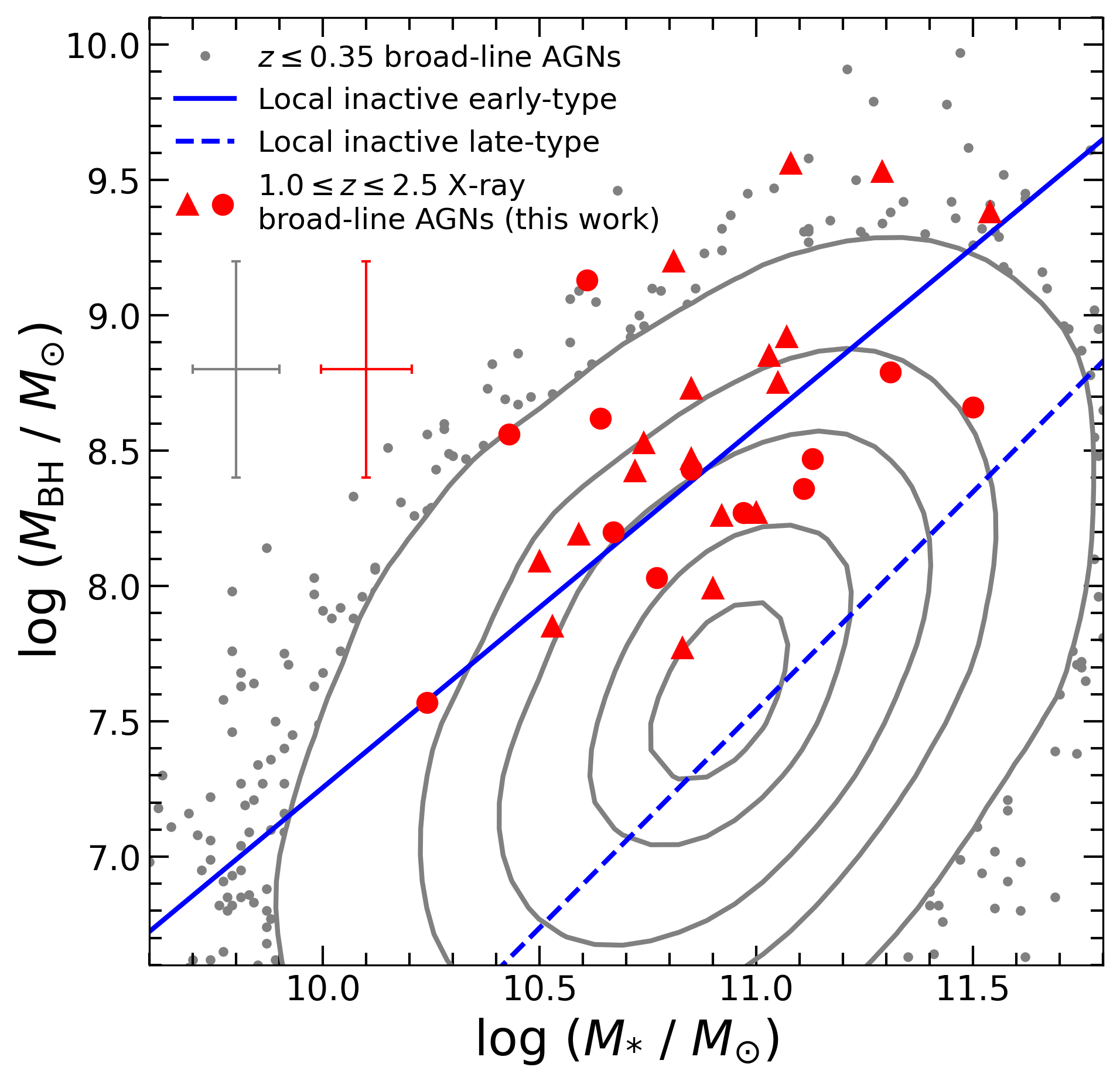}
\caption{The relation between $M_{\rm BH}$ and $M_*$ relation for $1.0\leq z \leq 2.5$ X-ray broad-line AGNs (red triangles: $z<1.8$; red dots: $z\geq1.8$). Gray contours and dots indicate 10\%, 30\%, 60\%, 80\%, and 95\% of the entire distribution and individual objects outside the 95\% contour of $z\leq0.35$ broad-line AGNs from \citet{Zhuang&Ho2023}, respectively. Solid and dashed blue lines represent best-fit relations for local early-type inactive galaxies and late-type inactive galaxies from \citet{Greene+2020ARA&A}, respectively. Errorbar indicates typical uncertainty.}
\label{fig13}
\end{figure}

\subsection{The BH Mass--Host Stellar Mass Relation}

\cite{Suh+2020ApJ} and \cite{Ding+2020ApJ} present $M_{\rm BH}$ measurements for their samples using single epoch spectra. As no line width and luminosity are provided in \cite{Suh+2020ApJ}, we adopt their $M_{\rm BH}$ calibration and convert line measurements in \cite{Ding+2020ApJ} to $M_{\rm BH}$ for consistence. In total, we have 30 objects at $1.0<z<2.5$ (median redshift 1.8). CID-452 is covered by both samples with reasonable agreement in $M_{\rm BH}$ between two measurements (0.28 dex difference). We adopt a typical uncertainty of 0.4 dex for $M_{\rm BH}$ estimated from single epoch method. 

Figure~\ref{fig13} shows $M_{\rm BH}$ versus $M_*$ relation for our $1.0<z<2.5$ X-ray broad-line AGNs. Taking at the face value, $1.0<z<2.5$ X-ray broad-line AGNs follow the relation defined by local inactive early-type galaxies but lie systematically above local inactive late-type galaxies \citep{Greene+2020ARA&A}. Comparing with nearby ($z\leq0.35$) optical broad-line AGNs from \citet{Zhuang&Ho2023}, $1.0<z<2.5$ AGNs are preferentially lying near the upper envelope of the distribution of low-$z$ AGNs. 

However, we caution that the position of AGNs on the $M_{\rm BH}-M_*$ plane can be significantly affected by various factors. These factors include the selection bias induced by the flux limit of the spectroscopic observations, the AGN duty-cycle dependent observational bias, and substantial measurement uncertainties in the mass terms \citep{Lauer+2007ApJ, Shen&Kelly2010ApJ, Schulze&Wisotzki2011A&A, Li2021mass}. As a result, both high-$z$ and local AGNs may be biased tracers of the intrinsic mass relation of the underlying SMBH population. Specifically, at high redshifts, the detection limit of the BH mass is limited by the depth of spectroscopic follow-up observations. Therefore, the observed mass relation is biased towards the most luminous and massive BHs. On the other hand, in the nearby universe, the AGN duty cycle tends to drop significantly toward massive BHs (as opposed to the roughly $M_{\rm BH}$-independent duty cycle at $z>1$; e.g., \citealt{Schulze2015}), making more massive BHs have smaller probability to be observed as AGNs. Consequently, nearby AGNs may preferentially populate the lower envelope of the $M_{\rm BH}-M_*$ plane \citep[e.g.,][]{Schulze&Wisotzki2011A&A, Li2021mass}. It is also possible that the mass relations of high-$z$ and local AGNs are intrinsically different due to evolution. For example, \cite{Zhuang&Ho2023} show that the position of AGNs on $M_{\rm BH}-M_*$ plane is connected with the properties of BH accretion and star formation. Considering more intense star formation and BH activities toward higher redshift, the $M_{\rm BH}-M_*$ relation may evolve with redshift. 

Moreover, we have not independently measured the broad-line AGN spectra to derive single-epoch BH masses using the latest recipes \citep{Shen2023}. The significant uncertainties associated with the single-epoch $M_{\rm BH}$ estimates ($\sim0.35$ dex for $\rm H\beta$ and Mg\,{\sc ii} and $\sim0.5$ dex for  C\,{\sc iv}; \citealt{Shen2023}), along with a potential systematic, luminosity-dependent bias resulted from uncorrelated variations between AGN luminosity and broad-line width, could significantly modify the intrinsic $M_{\rm BH}-M_*$ distribution by flattening its slope and increasing its scatter \citep{Shen2013, Li2021mass}. While conducting an in-depth analysis of these biases is beyond the scope of this paper, we plan to uniformly analyze all available spectroscopic data covering the COSMOS-Web footprint to measure broad lines and derive more reliable BH masses in future work. We will also perform a detailed modeling of selection biases and investigate the intrinsic $M_{\rm BH}-M_*$ relation and its cosmic evolution at $0.5\lesssim z \lesssim3.5$ using the most accurate BH mass and stellar mass measurements.

\section{Summary}\label{Sec7}

In this paper, we presented data reduction and PSF model construction procedures of JWST NIRCam imaging data in four filters (F115W, F150W, F277W, and F444W) from COSMOS-Web covering half of the entire footprint (0.28 deg$^2$), with average 5$\sigma$ point source depths of 27.4--28.3 mag. We selected a sample of 143 spectroscopically-confirmed $0.35\leq z \leq 3.5$ X-ray broad-line AGNs. Combining JWST NIRCam images with archival HST imaging data, we performed detailed multiwavelength simultaneous AGN-host image decomposition and SED fitting to AGN contamination subtracted-host photometries to obtain the structural and stellar properties of the host galaxies. Our main results are as follows: 

\begin{itemize}
\item{We find that PSFs in COSMOS-Web NIRCam mosaics have significant temporal and spatial variations in all four filters (F115W, F150W, F277W, and F444W). The temporal variation (standard deviation) of PSF FWHM decreases from $\sim2.8$\% (F115W) to 0.8\% (F444W) and is dominated by short timescale fluctuations. The spatial variation of PSF FWHM in individual observations are considerably larger compared to that of the temporal variation. PSF spatial variation in individual observations is dominated by random variations, with a standard deviation of $\gtrsim5$\% for F115W and F150W filters and $\sim2$\% for F277W and F444W filters.}

\item{We robustly detect stellar emission from the host galaxy in 142/143 AGNs. The median $f_{\rm host}$ is $\sim58$\% and $70$\% at rest-frame 5000\AA\ and 1 \micron, respectively.}

\item{Nearly 2/3 AGNs reside in star-forming galaxies based on the UVJ diagram. AGN hosts follow a similar stellar mass--size relation as non-AGN galaxies, with slightly smaller size for star-forming AGN hosts. Our results support the idea that a significant fraction of AGNs tends to be hosted by compact star-forming galaxies, which are associated with concentrated molecular gas and dust distribution as suggested by recent submm observations of AGNs at cosmic noon.}

\item{Non-axisymmetric features from stellar bar, spiral arms, and minor majors, are ubiquitous in X-ray broad-line AGNs. Together with a small major merger fraction (7--22\%) and the small peak \sersic\ index (1--2), our results emphasize the important role of secular processes and minor mergers in triggering AGN activities.}

\item{We do not find significant AGN-host offset in our sample, with median offset constrained to be as low as 200 pc. Clumpy host morphology of $z>1$ galaxies at rest-frame UV can lead to artificial offset measurements.}

\item{For a subsample of 30 AGNs with BH mass measurements from the literature, we find that they follow the $M_{\rm BH}-M_*$ relation of local inactive galaxies but are preferentially lying near the upper envelope of nearby AGNs. We caution that selection bias and intrinsic differences of AGNs at different redshifts may lead to the large, apparent discrepancies.}

\end{itemize}

{The reduced NIRCam imaging data from this work form the basis for our continued investigation on AGNs and their host galaxies in the COSMOS-Web field. We make publicly available all reduced NIRCam images and PSF models for broader applications of these JWST data at \url{https://ariel.astro.illinois.edu/cosmos\_web/}. We are working on additional data products, such as photometric catalogs and source classifications, and will make them publicly available in future works.}

\begin{acknowledgments}
We would like to thank the COSMOS-Web team for designing this project and making it possible to have this rich, public multi-wavelength data set that enables a broad range of community science. We thank Stefano Marchesi for providing us the X-ray spectral fitting catalog. Based on observations with the NASA/ESA/CSA James Webb Space Telescope obtained from the Barbara A. Mikulski Archive at the Space Telescope Science Institute, which is operated by the Association of Universities for Research in Astronomy, Incorporated, under NASA contract NAS5-03127. Support for Program numbers JWST-GO-02057 and JWST-AR-03038 was provided through a grant from the STScI under NASA contract NAS5-03127. 
\end{acknowledgments}

\vspace{5mm}
\facilities{JWST (NIRCam), HST (ACS)}

\software{
\texttt{astropy} \citep{2013A&A...558A..33A,2018AJ....156..123A}, 
\texttt{GALFITM} \citep{Haussler+2013MNRAS}, 
\texttt{jwst} \url{https://jwst-pipeline.readthedocs.io/en/latest/},
\texttt{Matplotlib} \citep{Hunter2007}, 
\texttt{Numpy} \citep{Harris2020}, 
\texttt{photutils} \citep{photutils}, 
\texttt{PSFEx} \citep{Bertin2011ASPC}, 
\texttt{scipy} \citep{scipy}, 
\texttt{SExtractor} \citep{1996A&AS..117..393B}
}

\appendix
\section{``Wisp'' and ``Claw'' Features}
\begin{figure*}[t]
\centering
\includegraphics[width=0.7\textwidth]{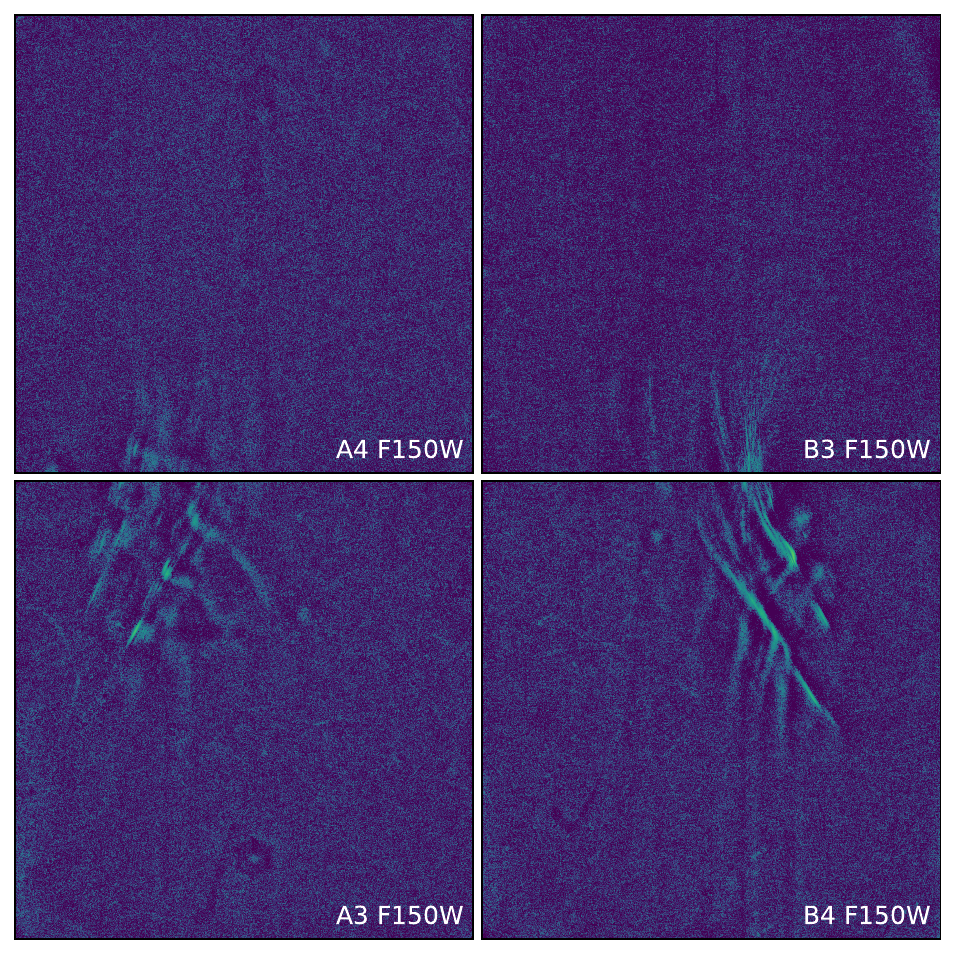}
\caption{Wisp templates in the F150W filter for COSMOS-Web observations created by median-stacking the source emission-masked images for four affected SW detectors: A4 (top left), A3 (bottom left), B3 (top right), and B4 (bottom right).}
\label{figa}
\end{figure*}

\begin{figure}[t]
\centering
\includegraphics[width=0.5\textwidth]{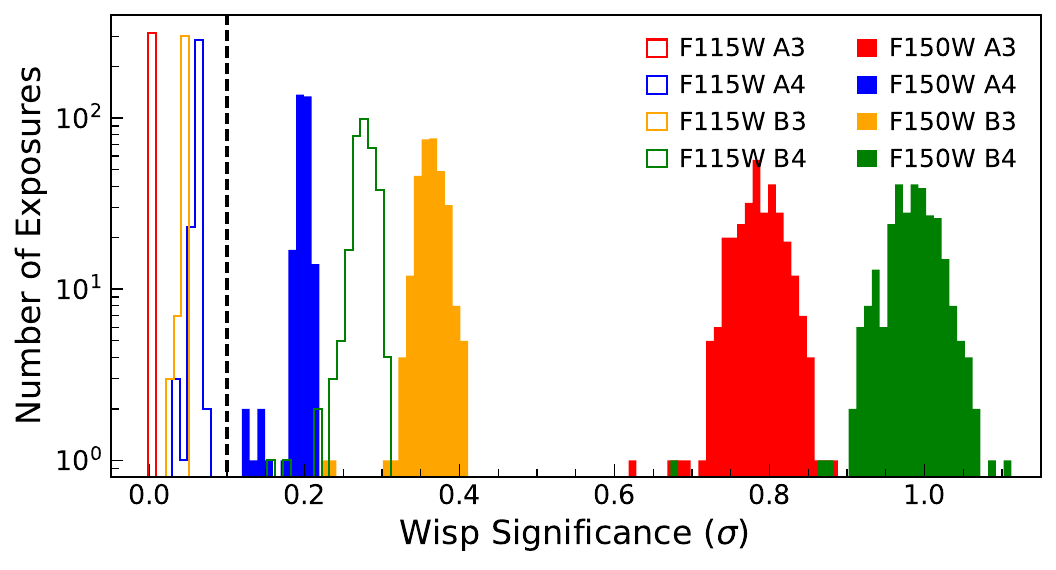}
\caption{Wisp significance in different detectors (A3: red, A4: blue, B3: orange, and B4: green) and filters (F115W: open step and F150W: filled step). Wisp significance is defined as the median ratio within a small window targeted at the brightest wisp feature between the wisps to be subtracted and the background noise ($\sigma$). Vertical dashed black line indicates our criterion ($>0.1\sigma$) for wisp subtraction: B4 of F115W and A3, A4, B3, and B4 of F150W. }
\label{figb}
\end{figure}

Wisp templates in the F150W filter are shown in Figure~\ref{figa}. {Wisps in the F115W filter have similar structures but much fainter brightness compared to those in the F150W filter. Figure~\ref{figb} presents the significance of wisp features in different detectors and filters. A3, A4, and B3 imaged with the F115W filter have wisp significance $<0.1\sigma$, suggesting negligible effects of wisps on these detectors. We only perform wisp subtraction in the four detectors imaged with the F150W filter and B4 imaged with the F115W filter.}

\begin{figure}[t]
\centering
\includegraphics[angle=90, width=0.6\textwidth]{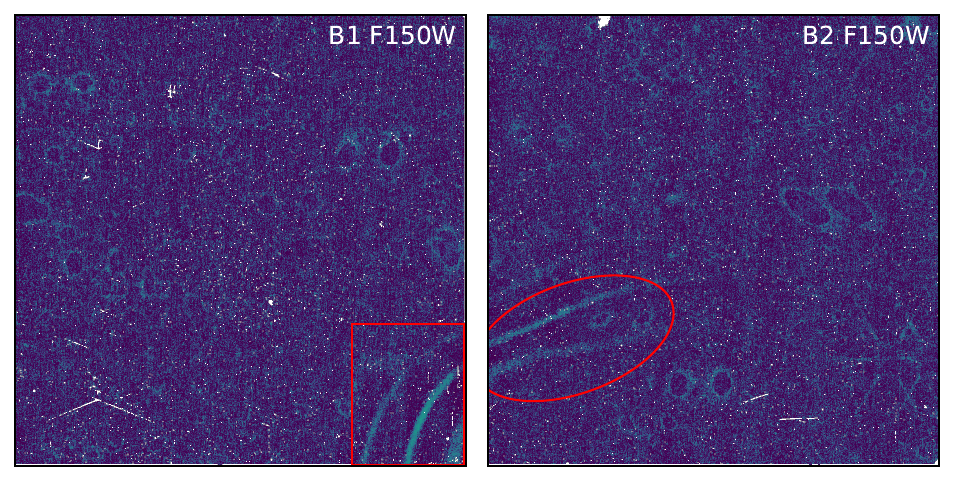}
\caption{An example of ``Claw'' features observed in the B1 and B2 detectors {(rotated by 90\degree\ counter-clockwise)} with the F150W filter in the Observation 043 highlighted with a red rectangle and ellipse. Images are created by median-stacking four dither exposures after masking sources. }
\label{figc}
\end{figure}

Figure~\ref{figc} shows an example of claw features observed in the B1 and B2 detectors with the F150W filter. {The position and strength of claws vary from field to field, with typical strength at the level of $\sim0.8$ times the background noise $\sigma$ in individual exposures, comparable to that of wisps in A3 detector in the F150W filter. We manually mask the pixels affected by claws for each exposure.}

\bibliography{sample631}{}
\bibliographystyle{aasjournal}



\end{CJK*}
\end{document}